\documentclass[%
superscriptaddress,
preprint,
nofootinbib,
amsmath,amssymb,
aps,
]{revtex4-2}
\usepackage[utf8]{inputenc}
\usepackage{amsmath}

\usepackage{graphicx} 
\usepackage[colorlinks=true]{hyperref} 
\hypersetup{colorlinks,
	citecolor=blue
}
\usepackage[dvipsnames]{xcolor}
\usepackage[caption=false]{subfig}
\usepackage{booktabs}
\usepackage{tikz}
\usetikzlibrary{patterns}
\usetikzlibrary{plotmarks}
\usetikzlibrary{matrix}
\usetikzlibrary{calc,shapes.misc,shapes.geometric}
\newcommand\Mark[2][]{%
	\tikz[baseline=(a.base)]{
		\node[inner sep=0pt,outer sep=0pt](a){\phantom{#2}};  
		\node[draw,black,thick,inner sep=2pt, rectangle,text=black,overlay,#1]  {#2};%
	}
}
\usepackage{gensymb} 
\usepackage{braket}
\usepackage[normalem]{ulem}

\begin{document}

\title{
New model comparison for semi-inclusive charged-current electron and muon neutrino scattering by $^{40}$Ar in the energy range of the MicroBooNE experiment}


\author{J. M. Franco-Patino}
\affiliation{
	Departamento de Física Atómica, Molecular y Nuclear, Universidad de Sevilla, 41080 Sevilla, Spain
}
\affiliation{
	Dipartimento di Fisica, Università di Torino, Via P. Giuria 1, 10125 Turin, Italy
}
\affiliation{
	INFN, Sezione di Torino, Via Pietro Giuria 1,10125, Turin, Italy
}

\author{S. Dolan }
\affiliation{CERN, European Organization for Nuclear Research, CH-1211 Geneva, Switzerland }

\author{R. Gonz\'alez-Jim\'enez}
\affiliation{
	Grupo de Física Nuclear, Departamento de Estructura de la Materia, Física Térmica y Electrónica and IPARCOS, Facultad de Ciencias Físicas, Universidad Complutense de Madrid, CEI Moncloa, Madrid 28040, Spain
}

\author{M. B. Barbaro}
\affiliation{
	Dipartimento di Fisica, Università di Torino, Via P. Giuria 1, 10125 Turin, Italy
}
\affiliation{
	INFN, Sezione di Torino, Via Pietro Giuria 1,10125, Turin, Italy
}

\author{J. A. Caballero}
\affiliation{
	Departamento de Física Atómica, Molecular y Nuclear, Universidad de Sevilla, 41080 Sevilla, Spain
}
\affiliation{
	Instituto de Física Teórica y Computacional Carlos I, Granada 18071, Spain
}

\author{G. D. Megias}
\affiliation{
	Departamento de Física Atómica, Molecular y Nuclear, Universidad de Sevilla, 41080 Sevilla, Spain
}

\date{\today}

\begin{abstract}
In this work we present a comparison of semi-inclusive muon and electron neutrino cross sections with $^{40}$Ar target measured by the MicroBooNE Collaboration with the predictions of an unfactorized model based on the relativistic distorted wave impulse approximation (RDWIA) and the SuSAv2-MEC model implemented in the neutrino event generator GENIE. The predictions based on the RDWIA approach, with a realistic description of the initial state and a phenomenological relativistic complex optical potential for the description of final state interactions, better describe the measured cross sections than GENIE-SuSAv2 and RDWIA with a purely real potential.
\end{abstract}

\maketitle

\section{Introduction}
Neutrino-nucleus interactions are one of the key inputs to measurements of neutrino oscillation parameters~\cite{ALVAREZRUSO20181}. Uncertainties associated with nuclear modeling are an important source of systematic error in both current neutrino oscillation experiments, NOvA~\cite{PhysRevLett.123.151803} and T2K~\cite{nature}, and future experiments such as DUNE~\cite{https://doi.org/10.48550/arxiv.2002.03005} and Hyper-Kamiokande~\cite{10.1093/ptep/ptv061}. Many present programs, like SBN~\cite{https://doi.org/10.48550/arxiv.1503.01520}, and future experiments, including DUNE~\cite{https://doi.org/10.48550/arxiv.1807.10334, https://doi.org/10.48550/arxiv.1807.10327, https://doi.org/10.48550/arxiv.1807.10340}, will employ liquid argon time projection chamber (LArTPC) detectors. As a consequence, neutrino-argon cross-section measurements have great importance, even more so considering that the main focus of neutrino-nucleus measurements in the past has been lighter nuclei like $^{12}$C and $^{16}$O.

The aim of accelerator-based neutrino oscillation experiments is to infer neutrino-oscillation parameters by comparing measured neutrino interaction event-rates at near and far detectors, usually as a function of a metric for neutrino energy reconstructed from final-state interaction products. In lower-energy experiments like T2K or MiniBooNE~\cite{PhysRevD.103.052002}, charged-current quasielastic (CCQE) scattering of the neutrino contributes a dominant interaction channel. Whilst different methods of neutrino energy reconstruction are used for different experiments, their spread and bias are usually driven by nuclear effects and non-CCQE contributions to measured CC0$\pi$ event samples. These nuclear effects include initial-state physics and final-state interactions (FSI), while the non-CCQE contributions correspond to two-particle-two-hole (2p2h) interactions, where the neutrino interacts with a pair of bound nucleons that are highly correlated, and interactions that produce a pion that is absorbed inside the nuclear medium via FSI. 

LArTPC detectors like SBN and DUNE offer the possibility of detecting additional particles in the final state, which improves the reconstruction of the neutrino energy. This makes it possible to obtain measurements that are highly sensitive to nuclear effects relative to inclusive measurements where only the final state lepton is detected. The one proton knockout process, where a lepton and one proton are produced~\cite{PhysRevD.90.013014}, has been studied within the plane-wave impulse approximation (PWIA)~\cite{PhysRevC.100.044620, PhysRevC.102.064626, PhysRevD.104.073008} and includes FSI using the relativistic distorted wave impulse approximation (RDWIA)~\cite{Gonzalez-Jimenez:2021ohu,PhysRevD.106.113005,but1,but2}. The T2K~\cite{PhysRevD.98.032003} and MINER$\nu$A~\cite{PhysRevLett.121.022504, PhysRevD.101.092001} collaborations have published $\nu_\mu-$CC0$\pi$ cross section measurements on $^{12}$C with one muon and at least one proton in the final state (denoted CC0$\pi$Np). In Ref.~\cite{PhysRevD.106.113005} we analysed the measurements on $^{12}$C within the unfactorized RDWIA approach. In this work we extend the analysis to semi-inclusive MicroBooNE measurements on $^{40}$Ar with two different topologies: zero pions, one lepton and at least one proton (CC0$\pi$Np)~\cite{PhysRevD.102.112013,PhysRevD.106.L051102} and zero pions, one lepton and exactly one proton (CC0$\pi$1p)~\cite{PhysRevLett.125.201803} in the final state. These measurements were made using the Booster Neutrino Beamline at Fermilab, which extends to 7 GeV and peaks around 0.7 GeV. As we did in~\cite{PhysRevD.106.113005} for $^{12}$C, in this paper we will also compare the semi-inclusive $^{40}$Ar measurements with the predictions from the inclusive model SuSAv2~\cite{PhysRevC.90.035501, PhysRevD.94.013012, PhysRevC.99.042501}, based on superscaling~\cite{Amaro:2004bs}, that has been implemented in the neutrino event generator GENIE~\cite{andreopoulos2015genie, ANDREOPOULOS201087}.  Strictly speaking, this model is only capable of predicting inclusive cross sections as function of the leptonic variables. However, assuming a factorization of the leptonic vertex and the initial nuclear state, it is possible for an inclusive model implemented in a neutrino event generator to make predictions about the ejected proton kinematics~\cite{PhysRevD.101.033003, dolan2021implementation}. It is important to point out that in the plane wave approach, the cross section factorizes into a single-nucleon term that takes care of the interaction between the lepton and a nucleon in the target, and the spectral function that incorporates nuclear effects. Factorization is not preserved in the RDWIA model. The meson exchange current (MEC) contribution to the 2p2h channel (following~\cite{RUIZSIMO2016124,PhysRevD.91.073004, PhysRevD.94.093004}) and pion absorption (following~\cite{Berger:2007rq,Dytman:2021ohr}) contribution are calculated with GENIE and added to the quasielastic cross sections for comparison to the available cross-section measurements.
	
The paper is organized as follows: In Section~\ref{sec2} we summarize the basic formalism for semi-inclusive neutrino-nucleus processes emphasizing the ingredients of the RDWIA approach. We discuss the model for the initial state and the description of the final state interactions using different approaches. Section~\ref{sec3} contains a detailed analysis of the results obtained comparing the theoretical predictions and data for CC0$\pi$Np and CC0$\pi$1p topologies. We consider both muon and electron neutrino scattering processes on $^{40}$Ar. Finally, in Section~\ref{sec4} we draw our main conclusions.

\section{\label{sec2}Semi-inclusive Neutrino-nucleus quasielastic scattering within RDWIA}

Following previous publications~\cite{Gonzalez-Jimenez:2021ohu, PhysRevD.106.113005, nikolakopoulos2022benchmarking}, in this section we briefly summarize the formalism to describe the one proton knockout channel where an incoming neutrino with momentum $\mathbf{k}$ interacts with a nucleus $A$, and a lepton and a proton, with momenta $\mathbf{k'}$ and  $\mathbf{p_N}$, respectively, are produced. In the laboratory frame, the flux-averaged six-differential semi-inclusive cross section is given by: 
	
	\begin{widetext}
		\begin{align}\label{semi-inclusive}
		\left <\frac{d\sigma}{dk'd\Omega_{k'}dp_{N}d\Omega^{L}_{N}}\right >&=\frac{G_F^2\cos^2{\theta_c}k'^2p_N^2}{64\pi^5}\int dk\, \Phi(k)\frac{W_B}{E_Bf_{\text{rec}}}L_{\mu\nu}H^{\mu\nu} ,
		\end{align}
	\end{widetext}
where $\Omega_{k'}$ and $\Omega_{N}^L$ are, respectively, the solid angles of the final lepton and the ejected proton, $\Phi(k)$ is the neutrino energy distribution (flux). The residual system $B$ can be left in an excited state with invariant mass $W_B$ and total energy $E_B$. $L_{\mu\nu}$ and $H^{\mu\nu}$ are the leptonic and hadronic tensors, and $f_{\text{rec}}$ is the recoil factor. All the information about initial-state dynamics and FSI is contained inside the hadronic tensor $H^{\mu\nu}$, which is built as the bilinear product of the matrix elements of the nuclear current operator between the initial and final nuclear states~\cite{Gonzalez-Jimenez:2021ohu, PhysRevD.106.113005, nikolakopoulos2022benchmarking}. Assuming that the impulse approximation is valid, the initial neutrino interacts with only one neutron of the target that is knocked out and turned into a proton. Then the proton travels through the residual nucleus undergoing FSI until it exits the nucleus.

In RDWIA the initial nucleons are described by a relativistic bound-state wave function obtained within the relativistic mean field (RMF) approach~\cite{HOROWITZ1981503}. We use a continuous missing energy ($E_m$) profile, denoted  $\rho\left(E_m\right)$, where each of the seven independent-particle shell model (IPSM) states $\alpha$ are modeled as  Maxwell-Boltzmann distributions
\begin{align}
	\rho_\alpha\left(E_m\right) = &\frac{4S_\alpha}{\sqrt{\pi}\sigma_\alpha}\bigg(\frac{E_m - E_\alpha + \sigma_\alpha}{\sigma_\alpha}\bigg)^2\\\nonumber
	&\times\exp\bigg[-\bigg(\frac{E_m - E_\alpha - \sigma_\alpha}{\sigma_\alpha}\bigg)^2\bigg]
\end{align}
with $E_\alpha$ the position of the peak, $\sigma_\alpha$ the width and $S_\alpha$ the occupancy of the shell.

An additional $1s1/2$ shell, called background, is included to account for the correlated nucleons that are not in the IPSM states. This background is parametrized as follows~\cite{Gonzalez-Jimenez:2021ohu, PhysRevD.106.113005}:
\begin{align}
	B\left(E_m\right) = S_ba\exp(-bE_m)
\end{align}
if $E_m > 100$ MeV, and
\begin{align}
	B\left(E_m\right) = \frac{S_ba\exp(-100\,b)}{\exp\big[-(E_m-c)/w\big]+1}
\end{align}
if $20 < E_m < 100$ MeV. The parameters are $a = 0.031127$ MeV$^{-1}$, $S_b$ the background occupancy of $^{40}$Ar, $b = 0.011237$ MeV$^{-1}$, $c = 40$ MeV and $w = 5$ MeV. The parametrization used in this work, that corresponds to the 22 neutrons in $^{40}$Ar, is summarized in Table~\ref{table: argon}. In this work we analyse the semi-inclusive quasielastic reaction induced by a neutrino beam, therefore, we focus on the configuration of the initial-state neutrons. From a theoretical point of view, anti-neutrino induced quasielastic cross sections can also be described by the RDWIA formalism, provided the proton initial state (see e.g.~\cite{JeffersonLabHallA:2022cit}); however, this reaction requires the detection of neutrons which is not possible with the current LArTPC technology. 

\begin{figure}[!htbp]
			\centering
			\includegraphics[width=0.49\textwidth]{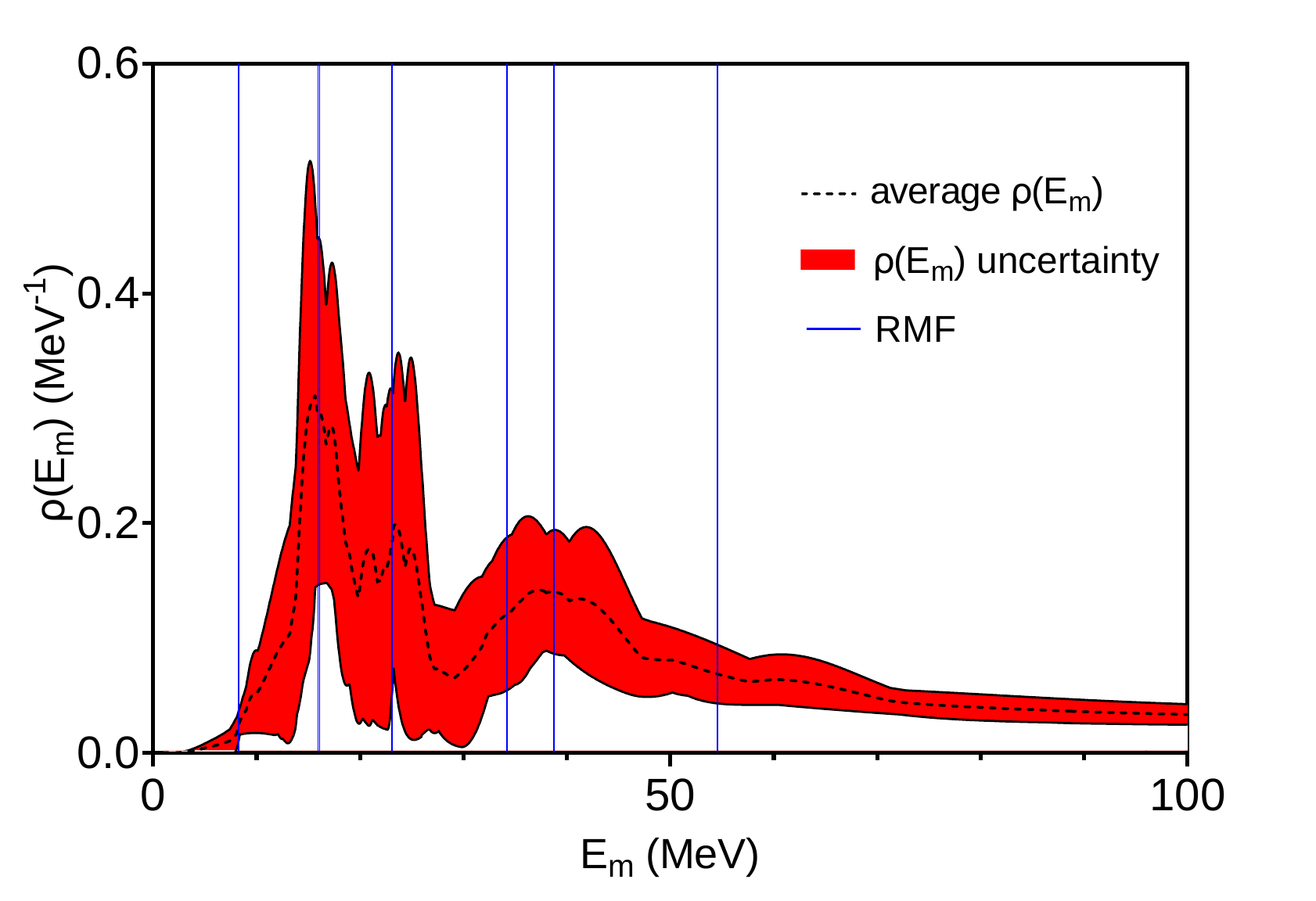}
			\caption{\label{rho_40ar} Missing energy profile of neutrons in $^{40}$Ar described by the parameterization given in Table~\ref{table: argon}. The red band corresponds to the uncertainties summarized in Table~\ref{table: argon}, the vertical blue lines show the positions of the RMF shells and the black dashed line shows the mean value of the distribution.}
		\end{figure}

In the next section, the RDWIA results are presented with colored bands that show an estimate of the error in the theoretical calculation due to the uncertainty on the modeling of the $^{40}$Ar missing energy profile. The uncertainties of the parameters that model this profile are shown in Table~\ref{table: argon}. The bands are constructed by randomly sampling the values of the missing energy profile parameters within their uncertainties with an uniform probability distribution. The number of neutrons in the background is such that the total number is 22, and the calculation is done only if the background contains between 15-25\% of the 22 neutrons, which is consistent with previous studies~\cite{ATTI20151,natureSRC,PhysRevLett.96.082501}. The modeled missing energy profile is shown in Fig.~\ref{rho_40ar} compared with the RMF predictions. 

Within the RDWIA approach FSI are included by solving the coupled differential equations derived from the Dirac equation with scalar and vector potentials that include the short-range strong interaction and the Coulomb potential. Regarding the potential that describes the strong interaction we consider two possibilities: a phenomenological complex relativistic optical potential (ROP) fitted to reproduce elastic proton-nucleus scattering data and the same RMF potential used to describe the initial state but multiplied by a phenomenological function that weakens the potential for increasing nucleon momenta~\cite{PhysRevC.100.045501, PhysRevC.101.015503} (denoted energy-dependent relativistic mean field or ED-RMF).
The parameterization of the ROP used in this work is the energy-dependent A-independent calcium
\footnote{To our knowledge, there is not any Dirac optical potential available for elastic proton$-^{40}$Ar scattering data, $^{40}$Ca is the closest nucleus for which they exist. We have seen that using any energy-dependent A-dependent potential also provided in~\cite{PhysRevC.47.297} produces similar results to the ones shown in this work.}
parameterization (EDAI-Ca)~\cite{PhysRevC.47.297}. The presence of an imaginary term in the ROP model leads to some flux loss as only the elastic scattering is described. On the contrary, the ejected nucleon wave functions in the ED-RMF model (pure real potential) are eigenstates of the same Hamiltonian used for the initial nucleon bound states. This ensures orthogonalization and Pauli blocking is incorporated consistently. Furthermore, the absence of the imaginary term in the potential ensures flux conservation, thus other channels in addition to the elastic one are incorporated. This explains why the ROP cross sections are significantly smaller than those obtained with the ED-RMF model. These predictions are also compared with the ones based on the SuSAv2 (inclusive) approach implemented in GENIE (see discussion above and \cite{PhysRevD.101.033003} for details).
	
	\begingroup
		\setlength{\tabcolsep}{6.3pt}
		\begin{table}[!t]
			\centering
			\begin{tabular}{cccccccc}
					\toprule\toprule
					$\alpha$ & &$E_\alpha$ (MeV) & &$\sigma_\alpha$ (MeV) & &$S_\alpha $ \\\midrule
					$1s_{1/2}$ & &55 $\pm$ 6 & & 30 $\pm$ 15 & &0.9 $\pm$ 0.15\\\midrule
					$1p_{3/2}$ & &39 $\pm$ 4& & 12 $\pm$ 6& &0.9 $\pm$ 0.15\\\midrule
					$1p_{1/2}$ & &34 $\pm$ 3& & 12 $\pm$ 6& &0.9 $\pm$ 0.15 \\\midrule
					$1d_{5/2}$ & &23 $\pm$ 2& &5 $\pm$ 3 & &0.75 $\pm$ 0.15 \\\midrule
					$2s_{1/2}$ & &16.1 $\pm$ 1.6& &5 $\pm$ 3 & &0.75 $\pm$ 0.15\\\midrule
					$1d_{3/2}$ & &16.0 $\pm$ 1.6& &5 $\pm$ 3 & &0.75 $\pm$ 0.15\\\midrule
					$1f_{7/2}$ & &9.869 $\pm$ 0.005& &5 $\pm$ 3 & &0.75 $\pm$ 0.15 \\
					\bottomrule\bottomrule
			\end{tabular}
			\caption{
   \label{table: argon} Parameterization of the missing energy distribution for the 22 neutrons in $^{40}$Ar used in this work. The shells are modeled as Maxwell-Boltzmann distributions (see the text). The spectroscopic factors $S_\alpha$ give the relative occupancy of the shell respect to the pure shell model occupancy. The background occupancy is fixed by the normalization condition $\sum_\alpha\int dE_m(2j_\alpha+1)\left[\rho_\alpha(E_m)+B(E_m)\right] = 22$, where $j_\alpha$ is the total angular momentum of the shell $\alpha$. The previous condition is fulfilled by setting $S_b = 3.05$, which results in around 4.3 neutrons in the background. The position of the $1f_{\frac{7}{2}}$ shell was set to the experimental neutron separation energy~\cite{Wang_2017}, and the others were set to the RMF values. The widths used in this model were inspired by those of the proton spectral function obtained from the analysis of the JLab $^{40}$Ar$\left(e,e'p\right)$ experimental data~\cite{JeffersonLabHallA:2022cit}.}
		\end{table}
	\endgroup
 
\section{\label{sec3}Results and discussion}
In this section we show MicroBooNE CC0$\pi$Np~\cite{PhysRevD.102.112013,PhysRevD.106.L051102} and CC0$\pi$1p~\cite{PhysRevLett.125.201803} measurements compared with the quasielastic predictions using the unfactorized RDWIA approach with different treatments of FSI and the implementation of the SuSAv2 model in GENIE~\cite{PhysRevD.101.033003}. The 2p2h-MEC (following~\cite{RUIZSIMO2016124,PhysRevD.91.073004, PhysRevD.94.093004}) and pion absorption (following~\cite{Berger:2007rq}) contributions are calculated with GENIE and added to the quasielastic cross sections for comparison to the available cross section measurements. Whilst a development version GENIE was used to produce these simulations, the results are identical to running GENIE \texttt{v3.2.0} in configuration \texttt{G21\_11b\_00\_000}. The processing of GENIE output and its comparison to experimental data was made using the NUISANCE framework~\cite{Stowell_2017}. Phase space restrictions applied for the comparison with the different experimental measurements are summarized in Table~\ref{kinematic_cuts}. A $\chi^2$-based analysis is presented in next section when discussing the results obtained for the cross sections.

	\begin{table*}[!t]
		\centering
		\resizebox{\textwidth}{!}{\begin{tabular}{cccccccccccccccccccccc}
				\toprule\toprule
				\Mark{1$\mu$CC0$\pi$Np}              &  &   $k'$  &   & & $\cos{\theta_l}$ & &    &  $p_N$  &  && $\cos{\theta_N^L}$&& &$\phi_N^L$& & & $\theta_{\mu p}$ &&&$\delta p_T$&\\\midrule
				               & & $> 0.1$ GeV & &  &      -   &  &  &0.3-1.2 GeV& &  &    -    & &&- & & & -&&&-&\\\midrule
				\Mark{1eCC0$\pi$Np}              &  &    &   & &  & &    &    &  && && && & &  &&&&\\\midrule
				               & & $> 30.5$ MeV & &  &      -   &  &  & $>$ 0.3105 GeV& &  &    -    & &&- & & & -&&&-&\\\midrule
				\Mark{1$\mu$CC0$\pi$1p}              &  &    &   & &  & &    &    &  && &\\\midrule
				               & & 0.1-1.5 GeV & &  &      $-0.65 < \cos{\theta_l} < 0.95$   &  &  &0.3-1.0 GeV& &  &    $> 0.15$    & &&145-215$^\circ$ & & & 35-145$^\circ$& & &$\delta p_T < 0.35$ GeV&\  \\ 
				\bottomrule\bottomrule
		\end{tabular}} \caption{\label{kinematic_cuts}Phase-space restrictions applied to $\nu_\mu$$-^{40}$Ar CC0$\pi$Np~\cite{PhysRevD.102.112013} and CC0$\pi$1p~\cite{PhysRevLett.125.201803} and $\nu_e$$-^{40}$Ar CC0$\pi$Np~\cite{PhysRevD.106.L051102} cross section measurements performed by MicroBooNE collaboration. The opening angle $\theta_{\mu p}$ is defined as the angle between the muon and the ejected proton and $\delta p_{T} =  \left| {\bf k'_T}+{\bf p_{N,T}}\right| $ is the transverse momentum imbalance~\cite{PhysRevC.94.015503} defined as the sum of the projections in the plane perpendicular to the neutrino direction of the muon and proton momenta. The index ``L" over the proton angles means they are defined in the laboratory frame (neutrino direction fixed in the $\hat{z}$ axis).} 
	\end{table*} 
 
\subsection{CC0$\pi$Np}
\begin{figure*}[!htbp]
		\centering 
		\begin{tikzpicture}
			\matrix[matrix of nodes]{
				\includegraphics[width=0.47\textwidth]{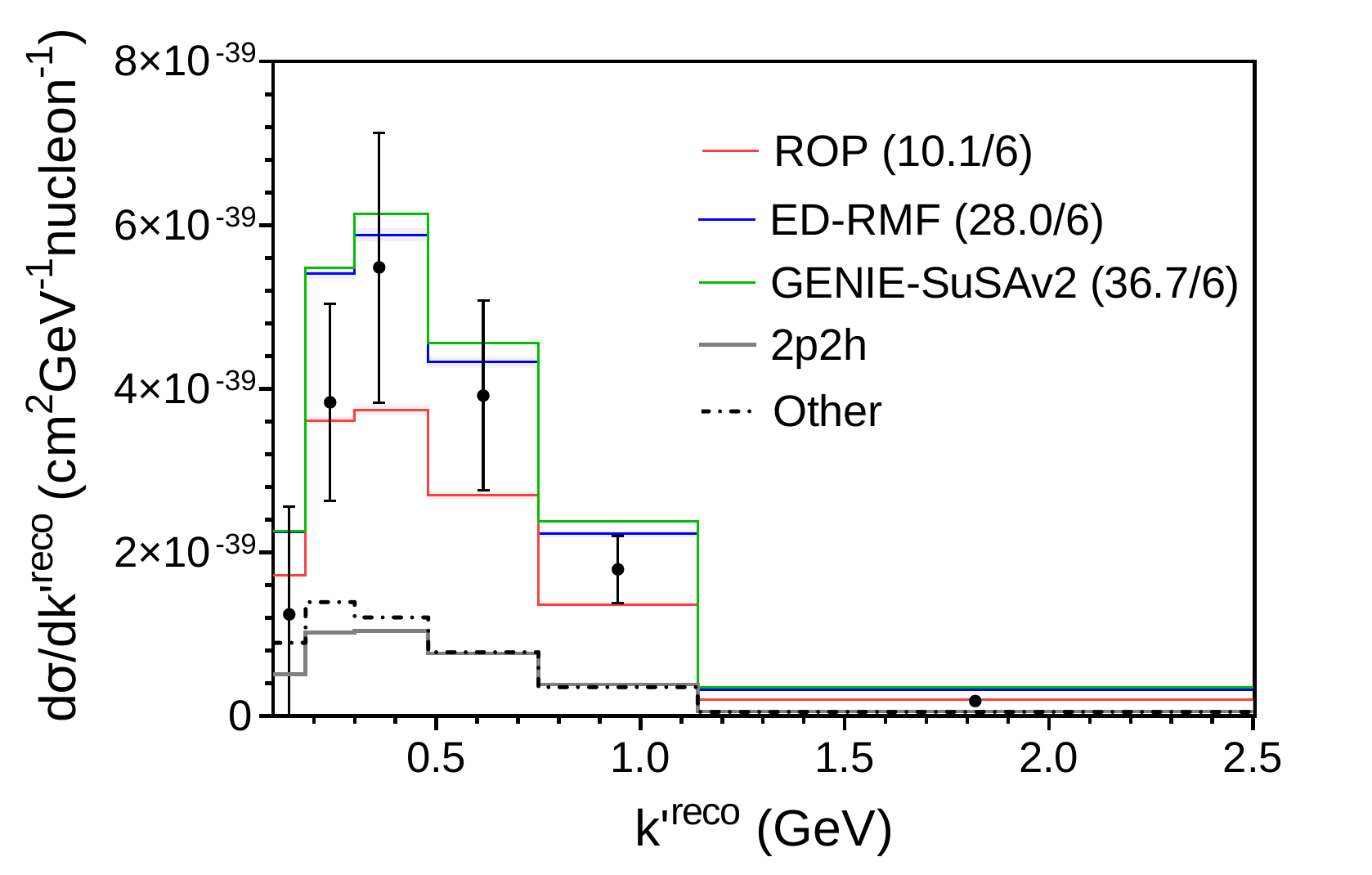}  \hspace{1.cm}
				\includegraphics[width=0.47\textwidth]{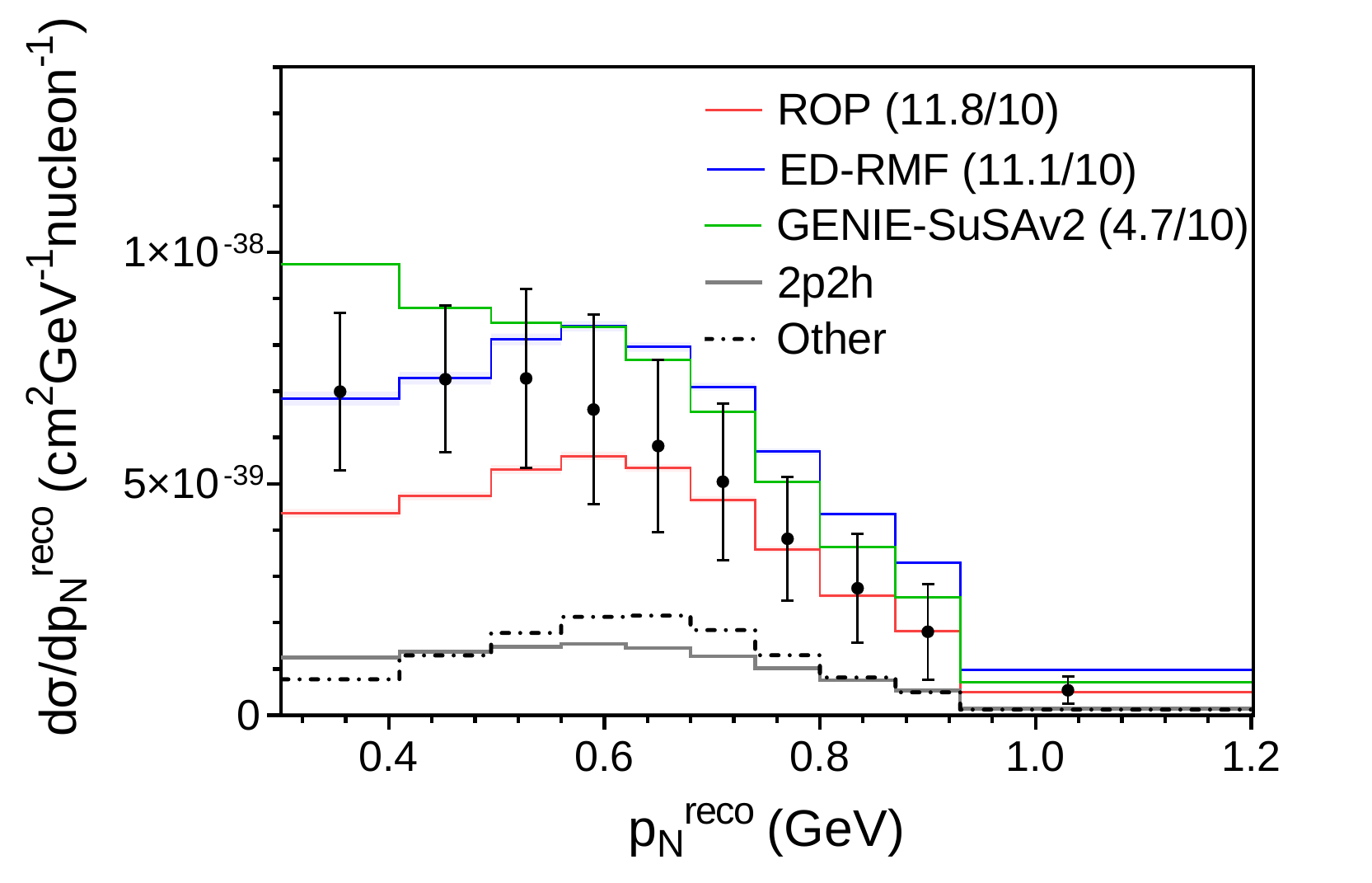} \\
				\includegraphics[width=0.47\textwidth]{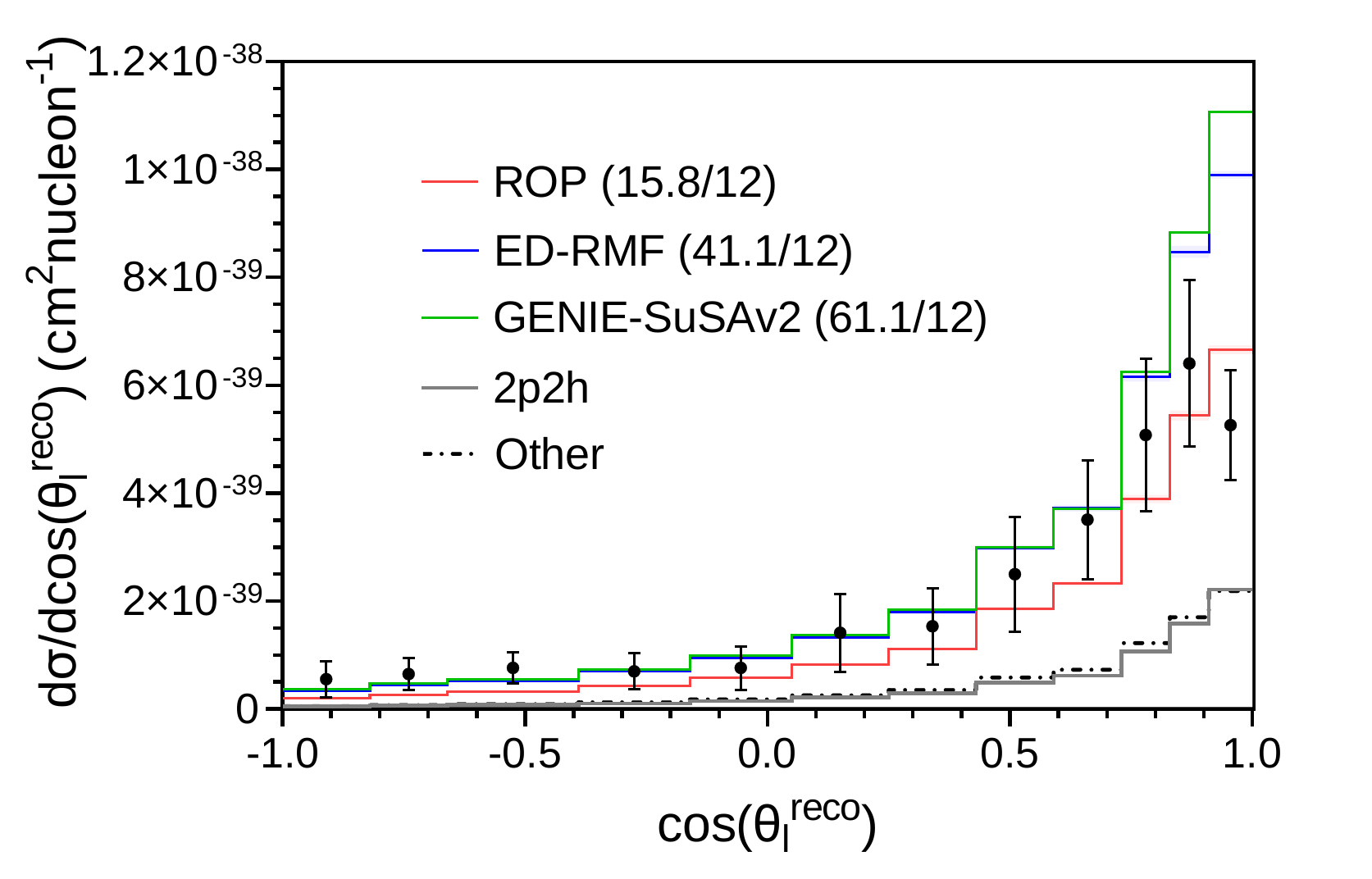}  \hspace{1.cm}
				\includegraphics[width=0.47\textwidth]{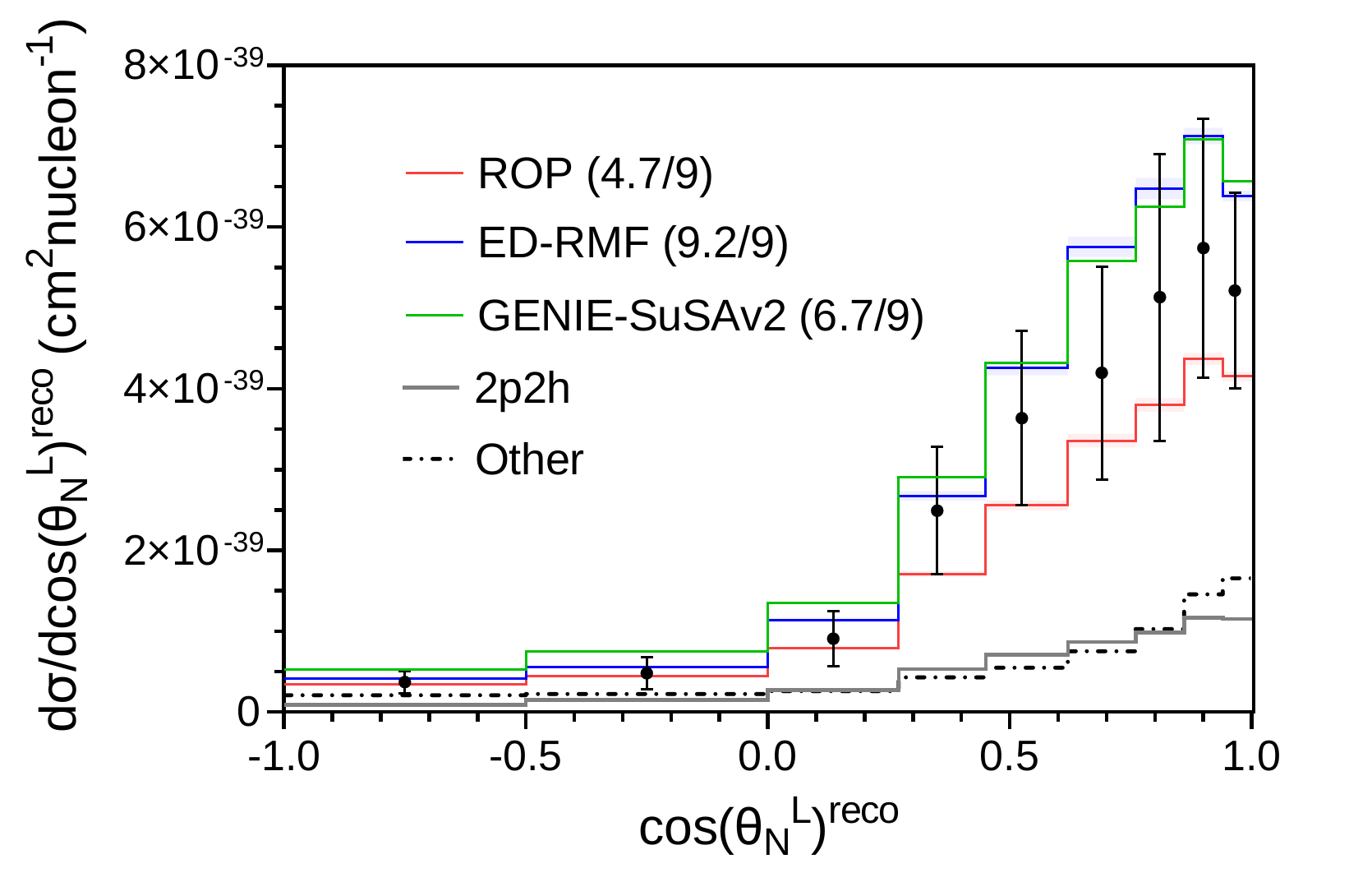} \\\vspace{2cm}
				\includegraphics[width=0.47\textwidth]{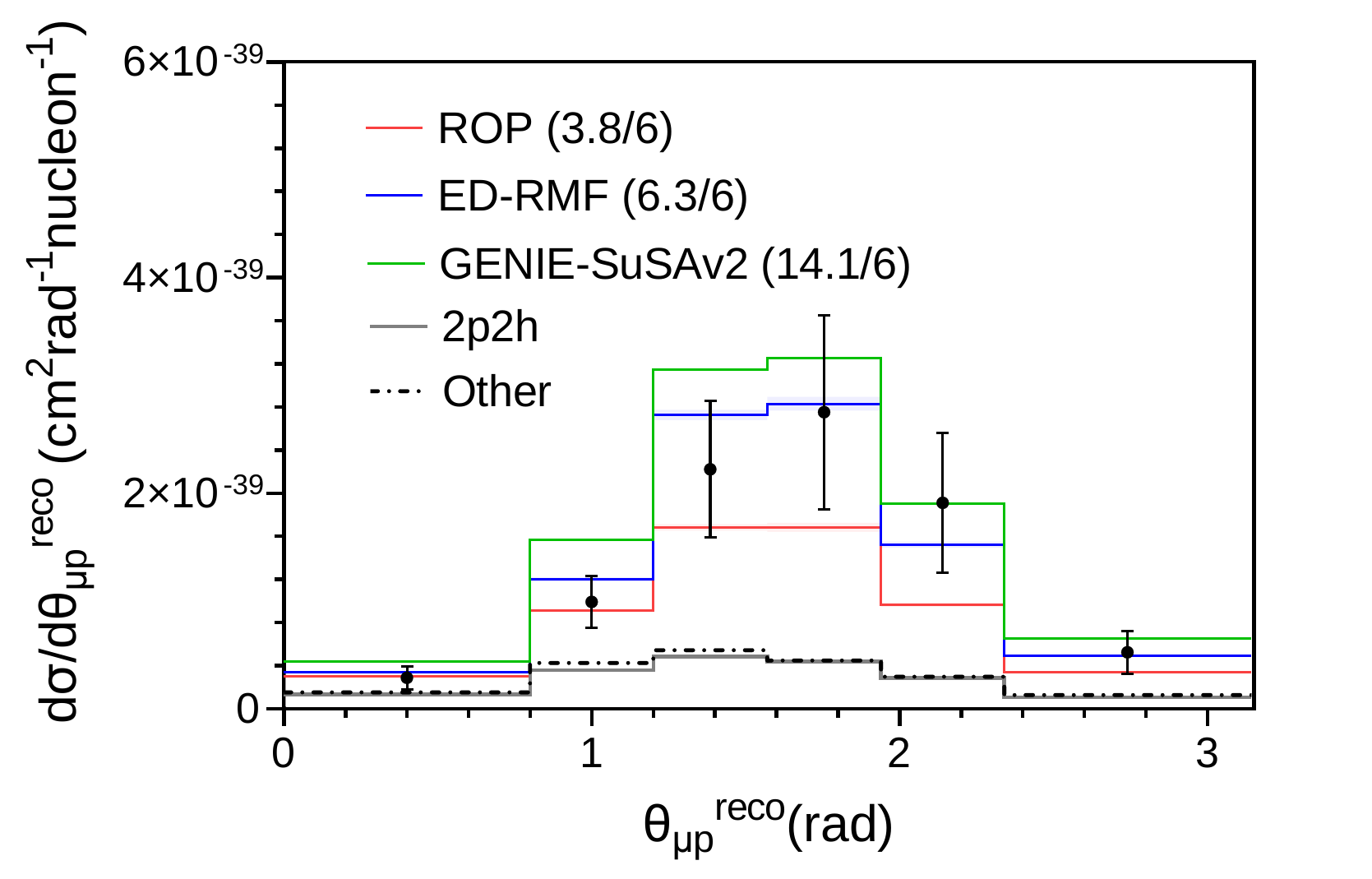} \\
			};
		\end{tikzpicture}
		\caption{\label{cc0piNp}MicroBooNE CC$0\pi$Np $\nu_\mu$$-^{40}$Ar cross sections as function of the reconstructed muon and proton momenta and scattering angles and the opening angle $\theta_{\mu p}^{\textnormal{reco}}$. All curves include the two-particle-two-hole (denoted 2p2h) and pion absorption (denoted other) contributions evaluated using GENIE (shown separately). Experimental results are from~\cite{PhysRevD.102.112013}. The bands, drawn for the ED-RMF and ROP models, represent the uncertainties associated with the modeling of the initial nuclear state. The $\chi^2/\textit{d.o.f.}$ ratio is given in brackets in the legend of each distribution.}
	\end{figure*}

In Fig.~\ref{cc0piNp} we compare the two RDWIA models previously described (ROP and ED-RMF) and the GENIE-SuSAv2 predictions with MicroBooNE 1$\mu$CC0$\pi$Np data for $^{40}$Ar~\cite{PhysRevD.102.112013}. The cross sections are shown as function of the muon and leading proton kinematics and also the opening angle $\theta_{\mu p}$. The experimental cross sections are given in terms of reconstructed variables, while our models predict the results as function of true variables. Therefore, we have applied the smearing matrix~\cite{PhysRevD.102.112013} to all the theoretical results shown in Fig.~\ref{cc0piNp}. Whilst the GENIE-SuSAv2 and ED-RMF models are in poor agreement with the measured $k'^{\textnormal{reco}}$ distribution in Fig.~\ref{cc0piNp}, the ROP model provides a reasonable description of it with a $\chi^2$ of $\sim$10 for 6 degrees of freedom (\textit{d.o.f.}). The shape of the $p_N^{\textnormal{reco}}$ distribution is correctly reproduced by the ED-RMF or ROP models once the 2p2h and other contributions are taken in account, although the ED-RMF model overestimates the measurements in the $0.65 < p_N^{\textnormal{reco}} < 0.9$ GeV range. It is interesting to note that the GENIE-SuSAv2 model overestimates the experimental measurement at very low $p_N^{\textnormal{reco}}$ (although its good agreement at large $p_N^{\textnormal{reco}}$ actually leads to a lower $\chi^2$).
	
The shape and magnitude of the $\cos{\theta_l}^{\textnormal{reco}}$ and $\cos{\theta_N^L}^{\textnormal{reco}}$ angular distributions (Fig.~\ref{cc0piNp}) are well described by all the models except at very forward muon scattering angles where all overestimate the cross-section measurement, although it is worth noting that this is less severe in the case of ROP, which provides a quantitatively good description of the distribution. Regarding the $\theta_{\mu p}^{\textnormal{reco}}$ distribution, ED-RMF appears to better describe the shape of the experimental measurement, but both ROP and ED-RMF are quantitatively compatible with data.
 
\begin{figure*}[!htbp]
		\centering 
		\begin{tikzpicture}
			\matrix[matrix of nodes]{
				\includegraphics[width=0.47\textwidth]{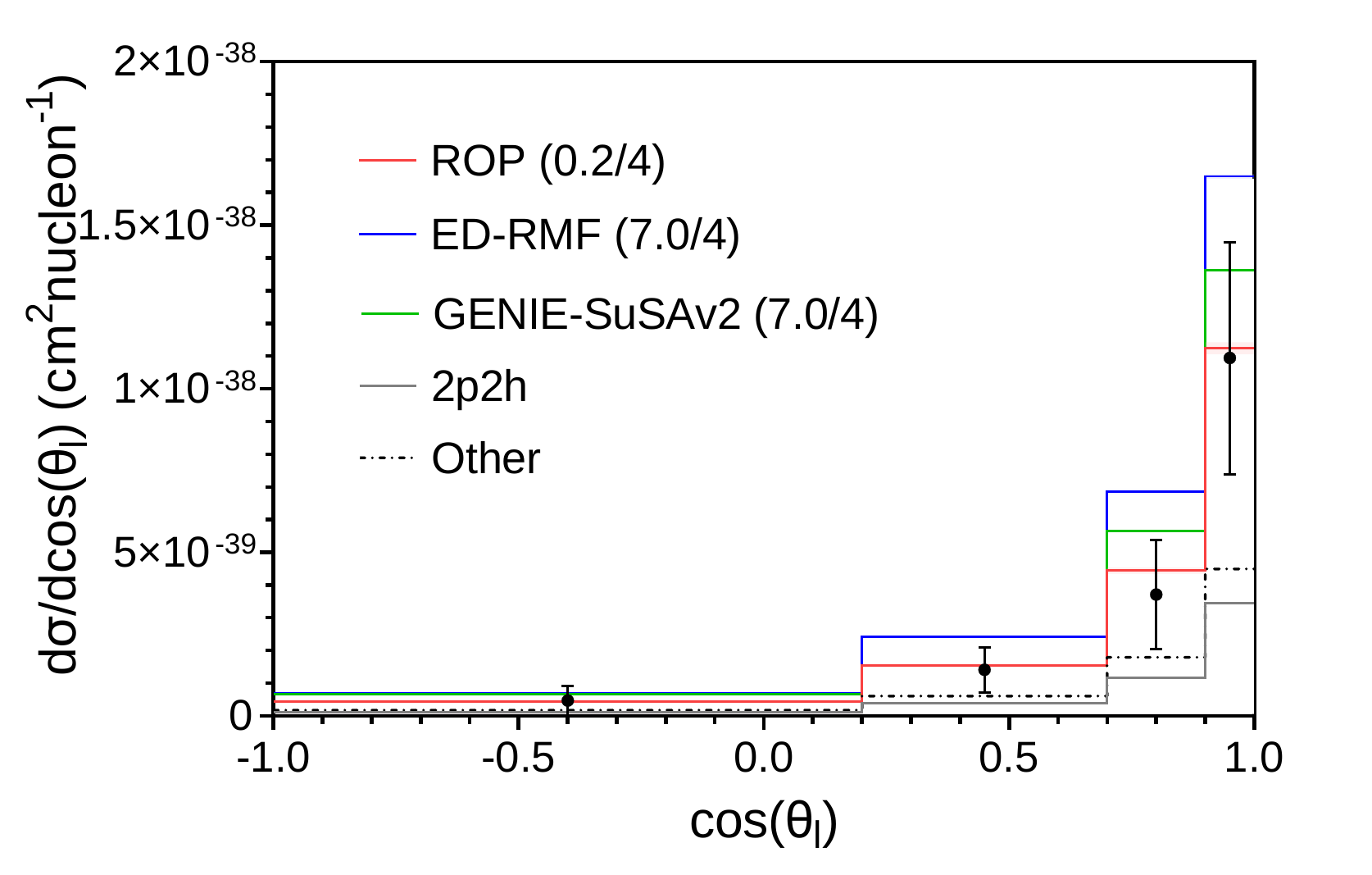}  \hspace{0.35cm}
				\includegraphics[width=0.47\textwidth]{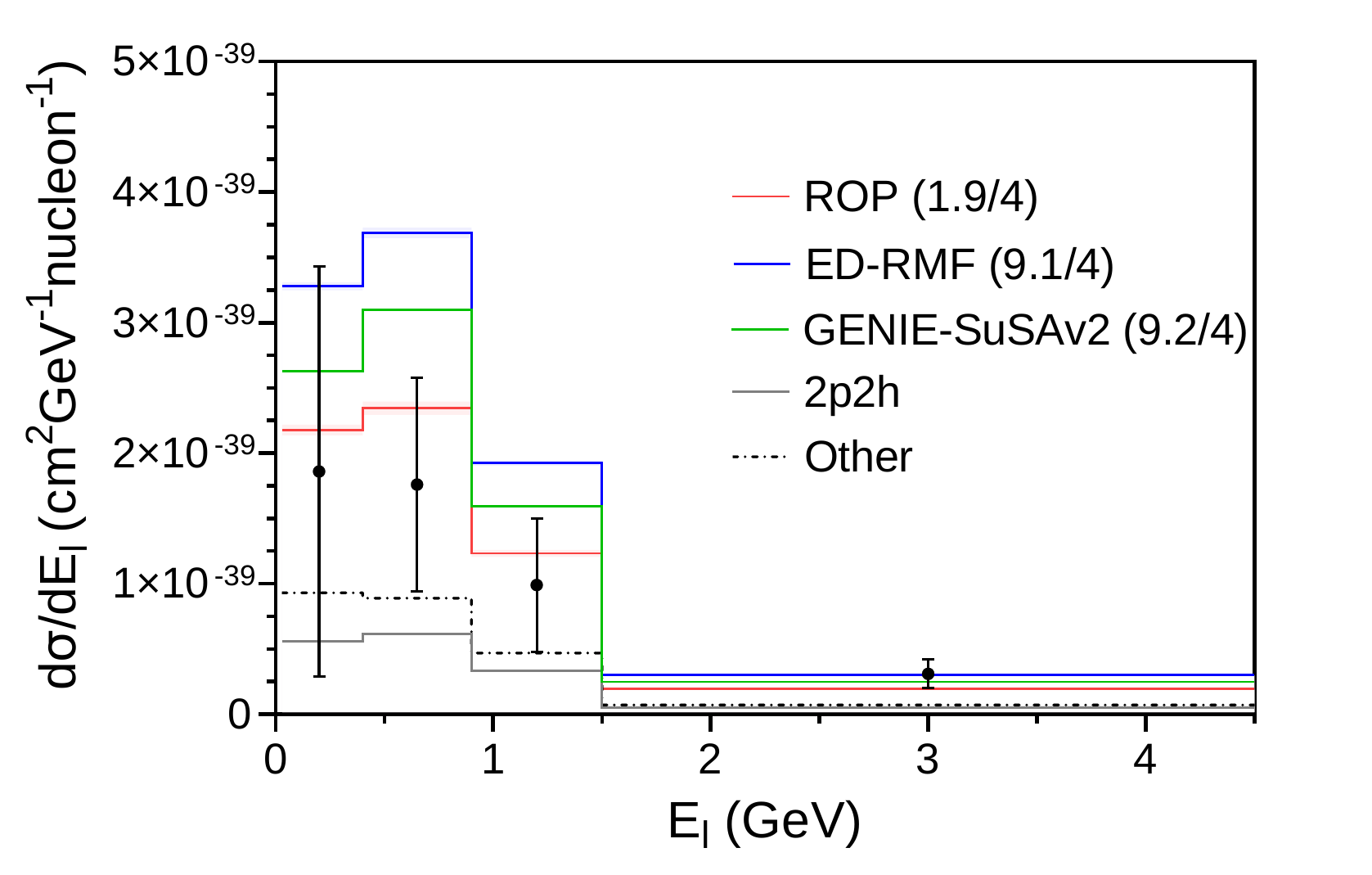} \\
				\includegraphics[width=0.47\textwidth]{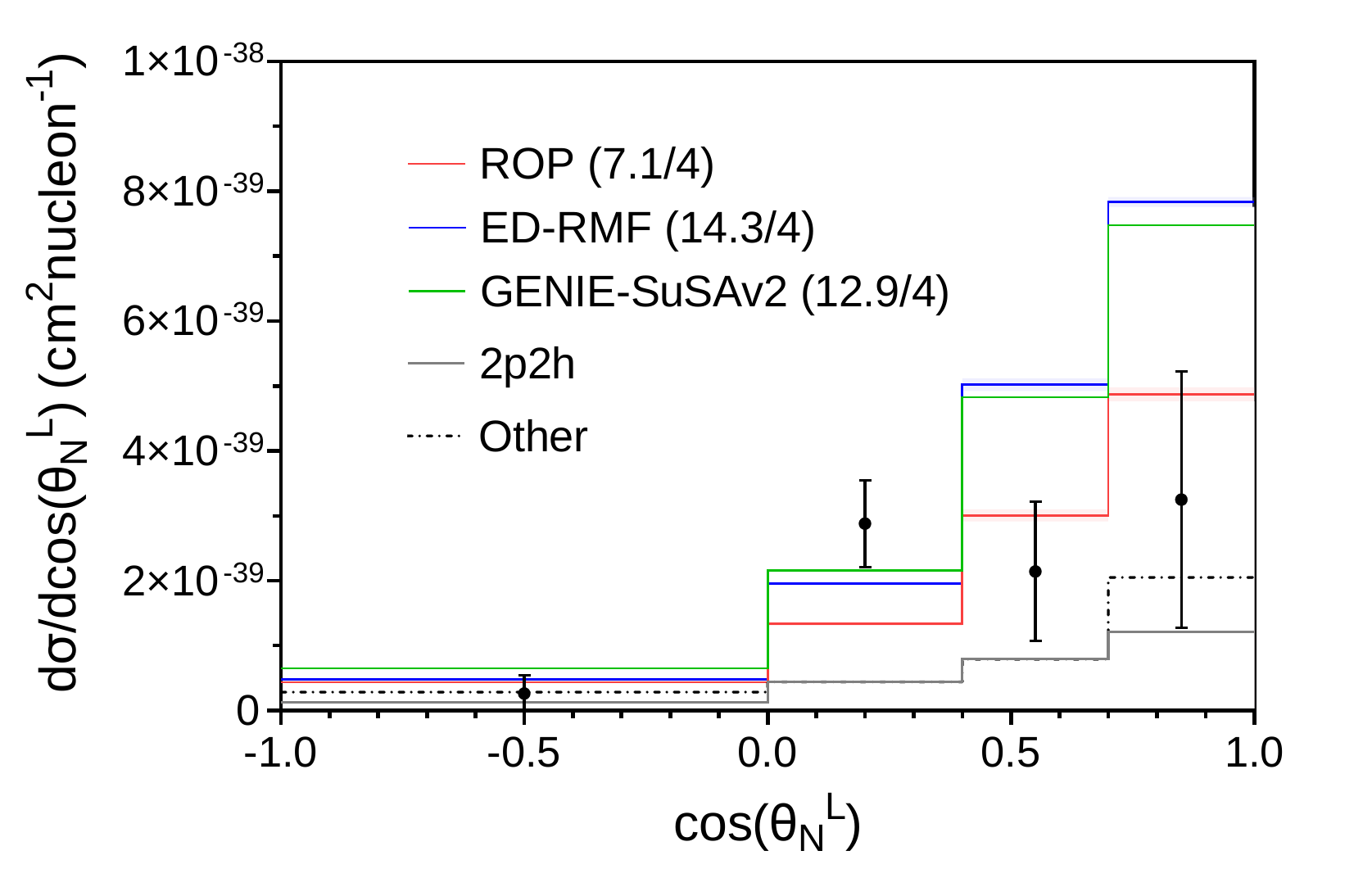}  \hspace{0.4cm}
				\includegraphics[width=0.47\textwidth]{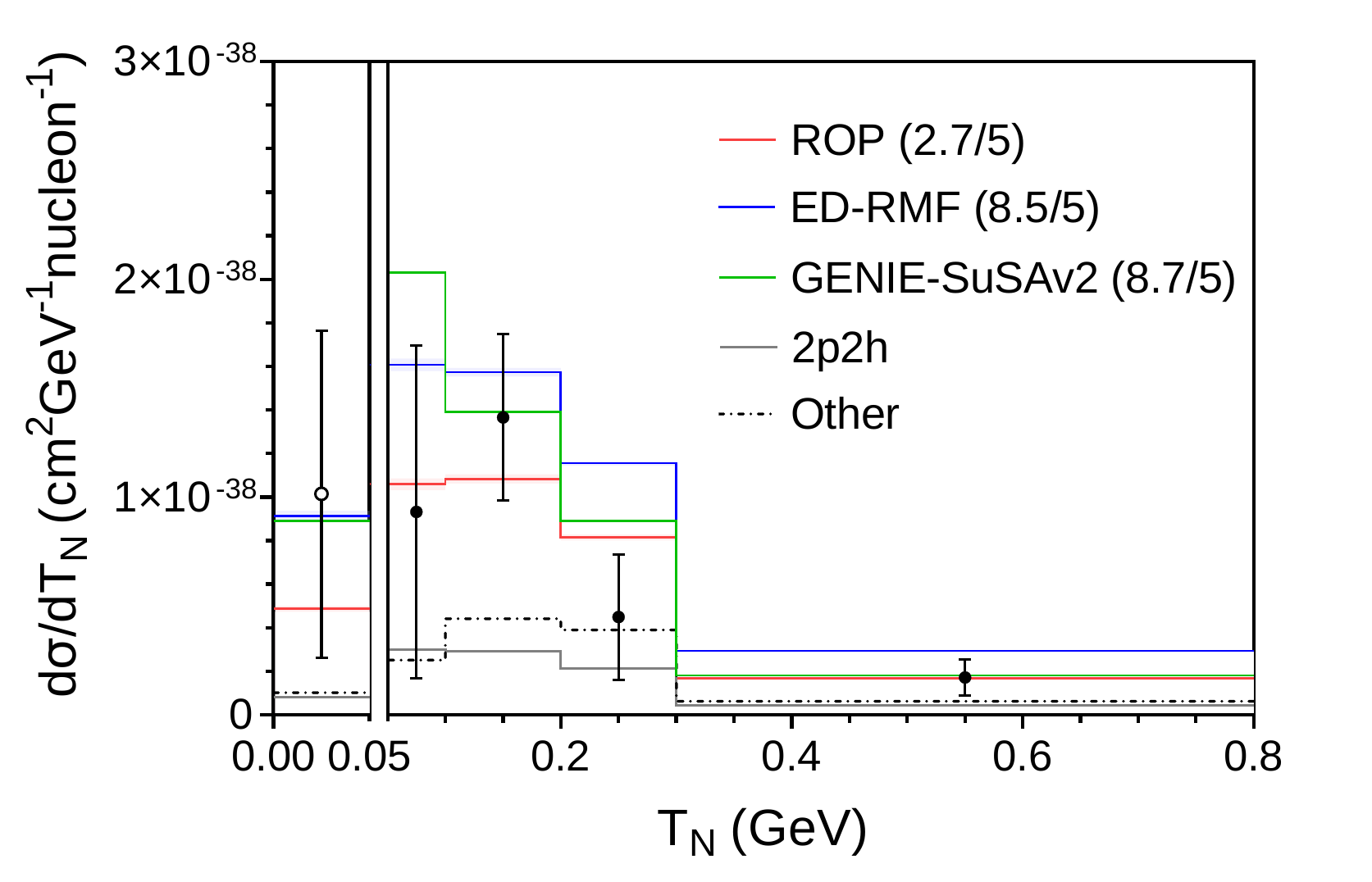} \\
			};
		\end{tikzpicture}
		\caption{\label{electron_cc0piNp}
		MicroBooNE CC$0\pi$Np $\nu_e$$-^{40}$Ar cross sections as function of the electron scattering angle and energy ($E_l$) and proton kinetic energy ($T_N$) and scattering angle. All curves include the two-particle-two-hole (denoted 2p2h) and pion absorption (denoted other) contributions evaluated using GENIE (shown separately). Experimental results are from~\cite{PhysRevD.106.L051102}. The bands, drawn for the ED-RMF and ROP models, represent the uncertainties associated with the modeling of the initial nuclear state. The single white point in the $T_N$ distribution between $0 < T_N < 50$ MeV corresponds to an extra 1e0p0$\pi$ (one electron, no protons above 50 MeV threshold and any number of protons with kinetic energy below the 50 MeV threshold) measurement performed by MicroBooNE~\cite{PhysRevD.106.L051102}. For this single point, additional phase space restrictions on the electron energy ($E_l > $ 0.5 GeV) and the electron scattering angle ($\cos{\theta_l} > $ 0.6) are applied. Note that for this bin, the GENIE SuSAv2 prediction is the same as the ED-RMF prediction and so cannot easily be seen on the plot. The $\chi^2/\textit{d.o.f.}$ ratio is given in brackets in the legend of each distribution.}
	\end{figure*}

In Fig.~\ref{electron_cc0piNp} we compare the different theoretical models with MicroBooNE 1eCC0$\pi$Np data on $^{40}$Ar~\cite{PhysRevD.106.L051102} as function of electron energy and scattering angle, and the final proton kinetic energy ($T_N$) and scattering angle. Additionally, for the $T_N$ distribution presented in Fig.~\ref{electron_cc0piNp}, MicroBooNE collaboration provides one extra data point (0 $< T_N <$ 50 MeV) that corresponds to events with one electron, no protons above $T_N$ = 50 MeV threshold and any number of protons below the threshold. Although the experimental measurements are statistically limited and the error bars are large, the ROP model seems to describe better all the measurements presented in Fig.~\ref{electron_cc0piNp}, whilst both GENIE-SuSAv2 and ED-RMF tend to overestimate them, although it should be noted that there is little quantitative power to statistically separate the models.
	
\subsection{CC0$\pi$1p}
	
\begin{figure*}[!htbp]
	\begin{tikzpicture}
		\matrix[matrix of nodes]{
			\includegraphics[width=0.32\textwidth]{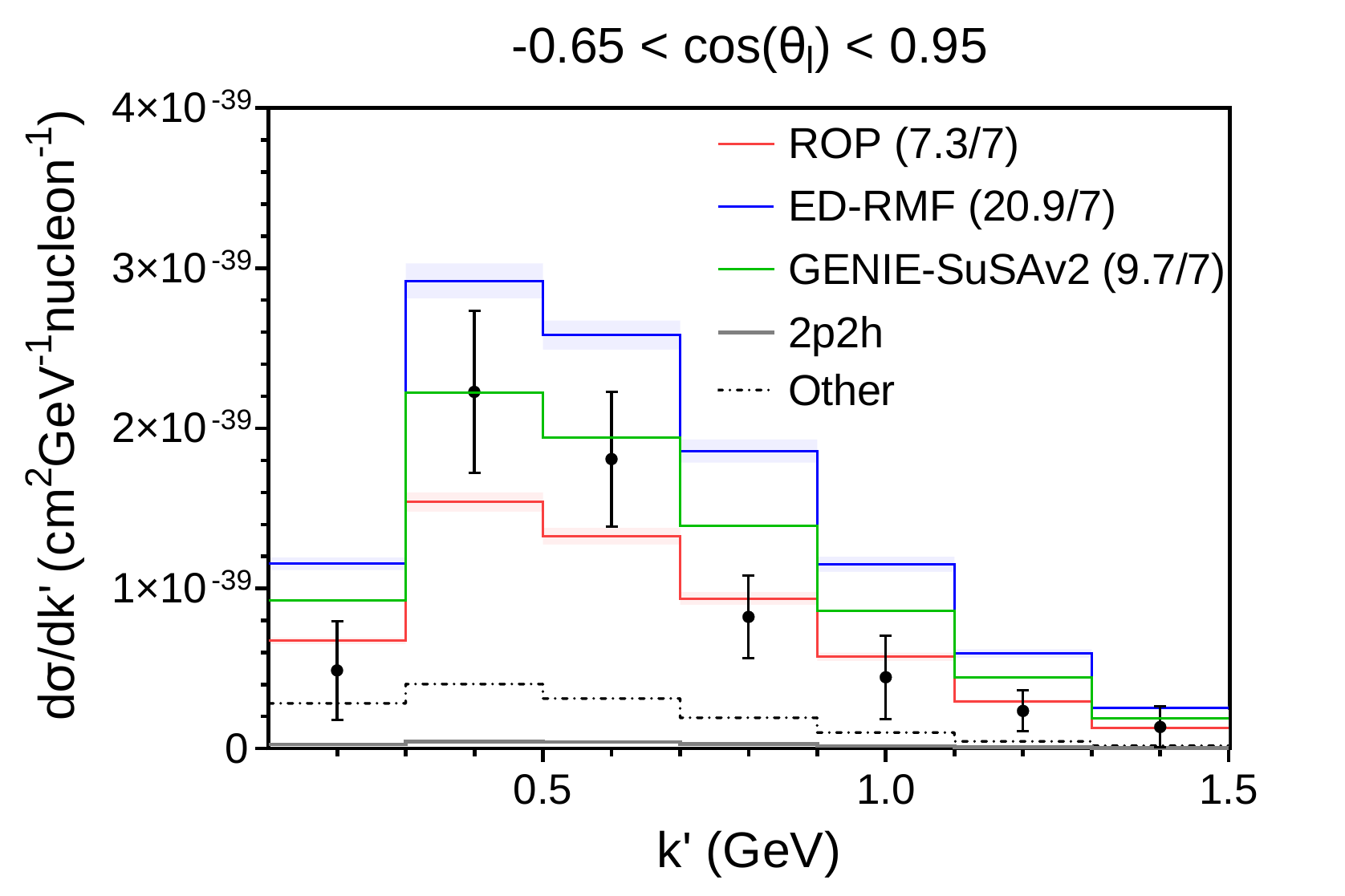}&\hspace{0.5cm}
			\includegraphics[width=0.32\textwidth]{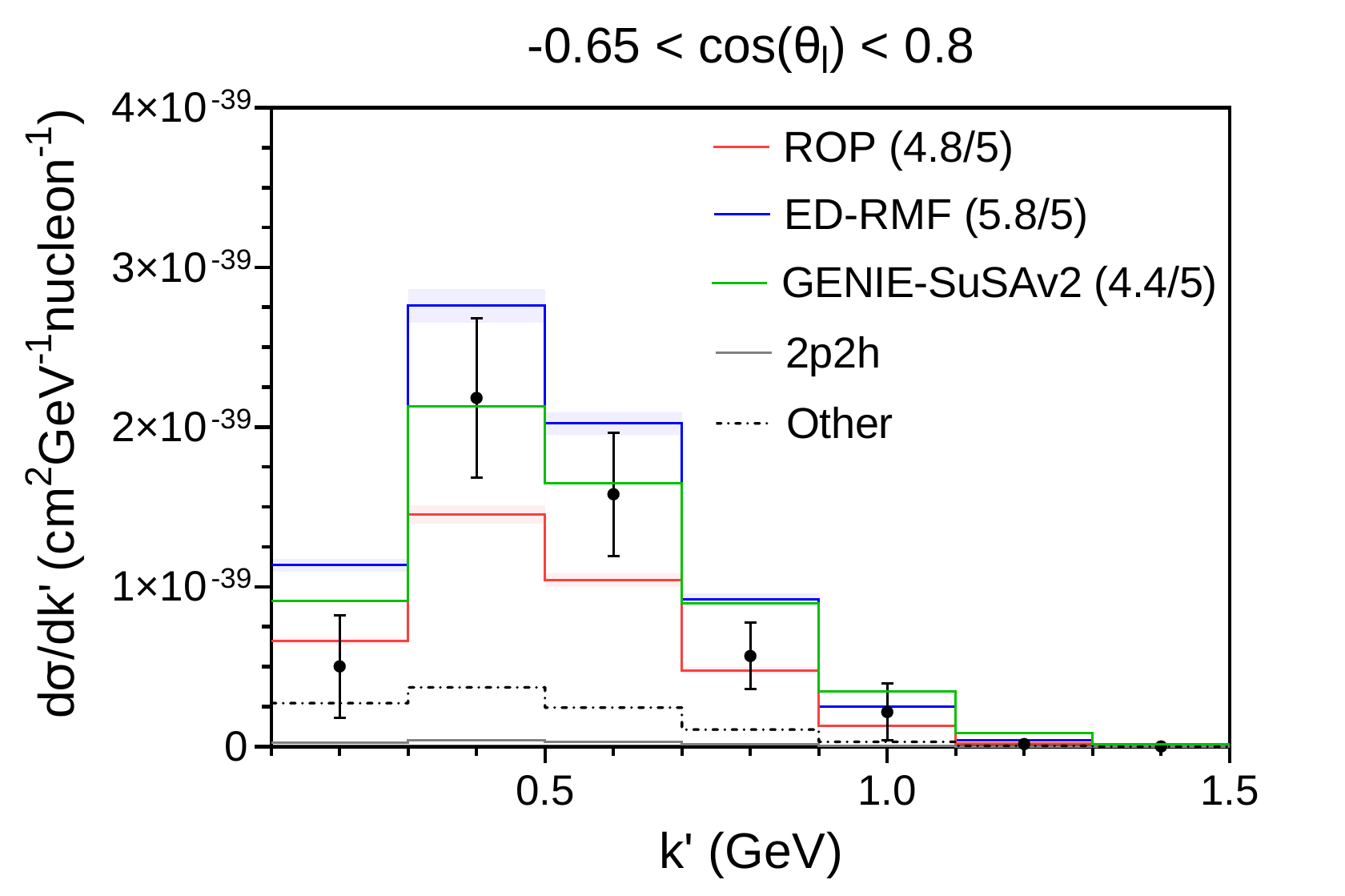}&\\ \includegraphics[width=0.32\textwidth]{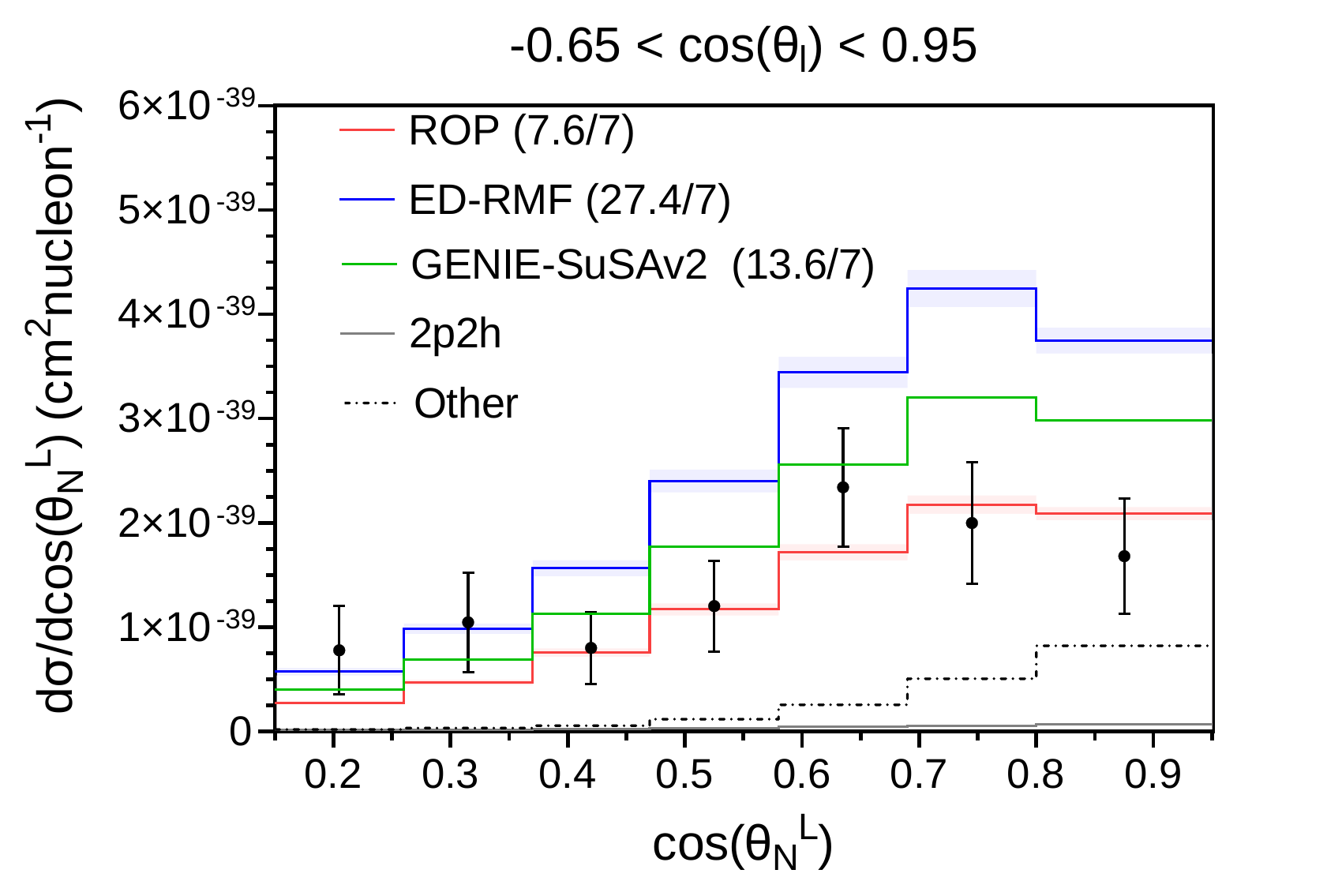} &\hspace{0.5cm}
			\includegraphics[width=0.32\textwidth]{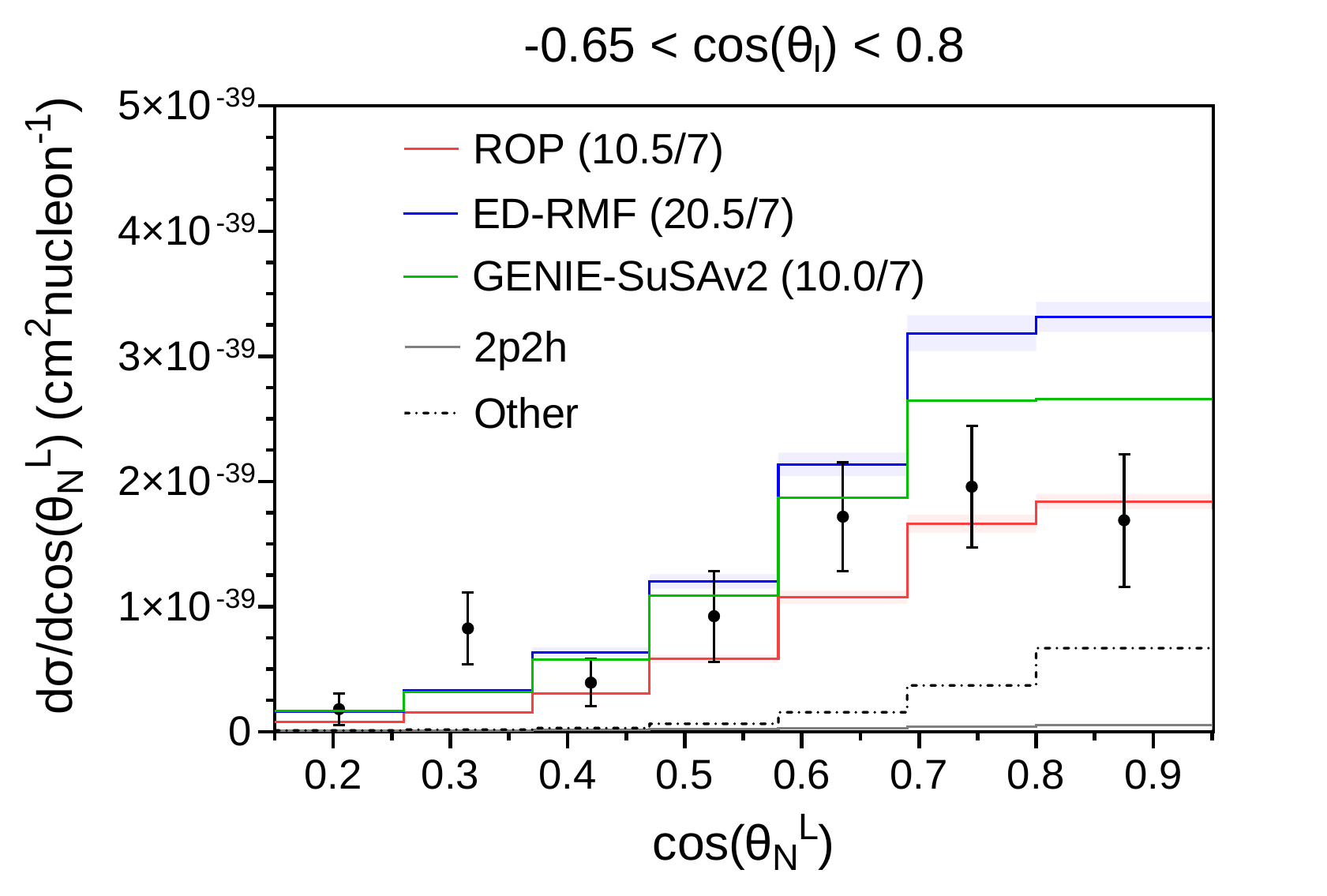} & \\
			\includegraphics[width=0.32\textwidth]{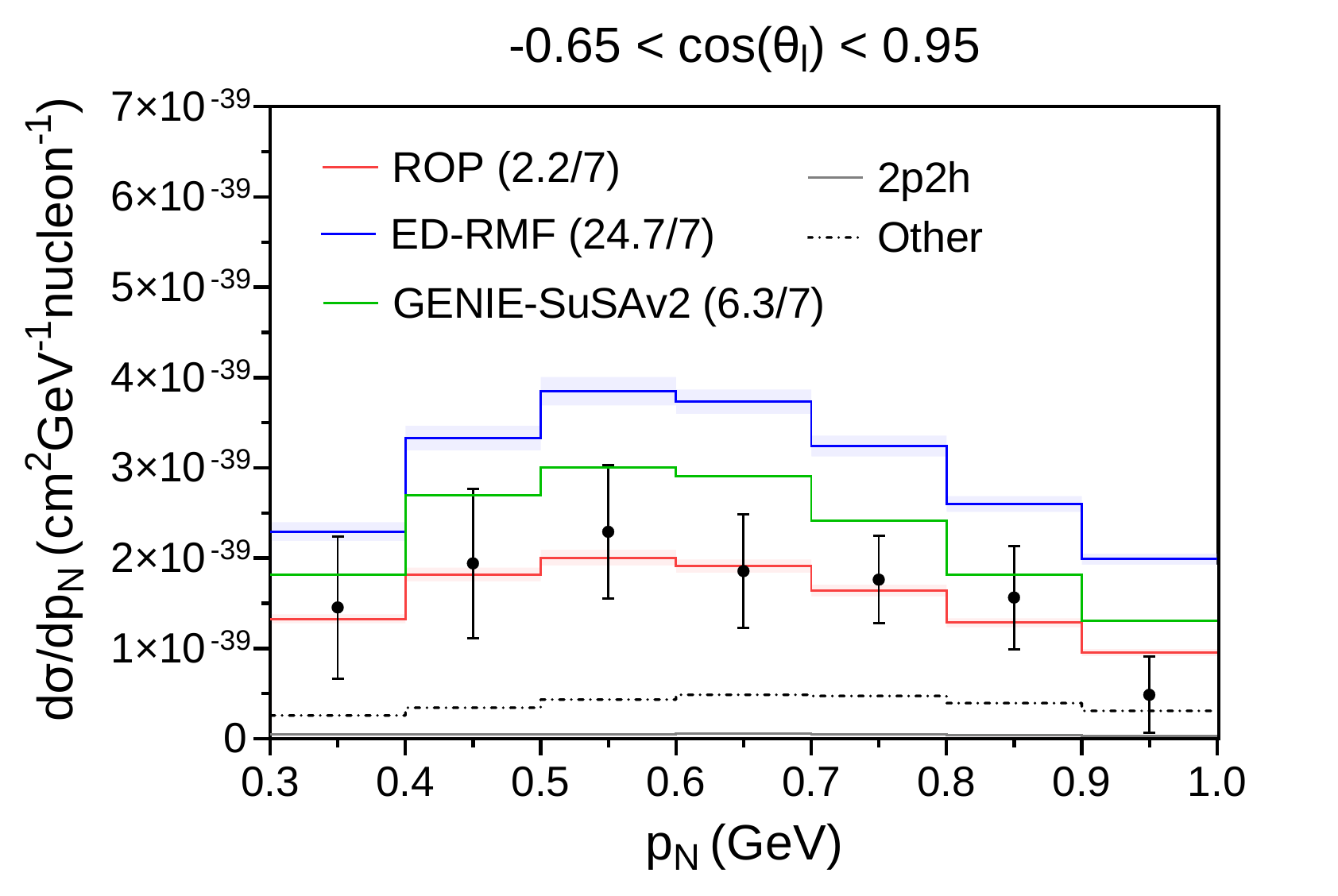} &\hspace{0.5cm}
			\includegraphics[width=0.32\textwidth]{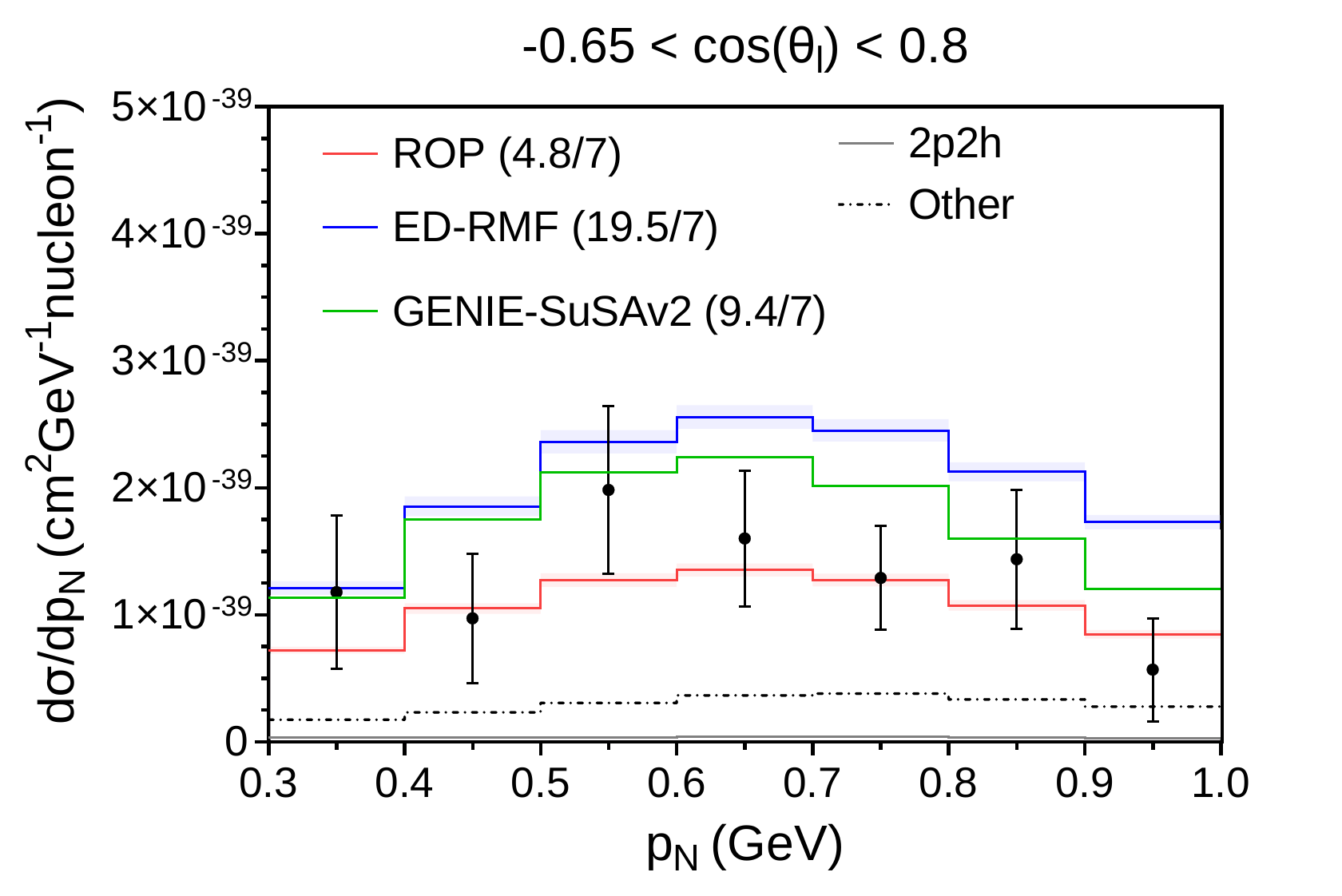} &\\ \includegraphics[width=0.32\textwidth]{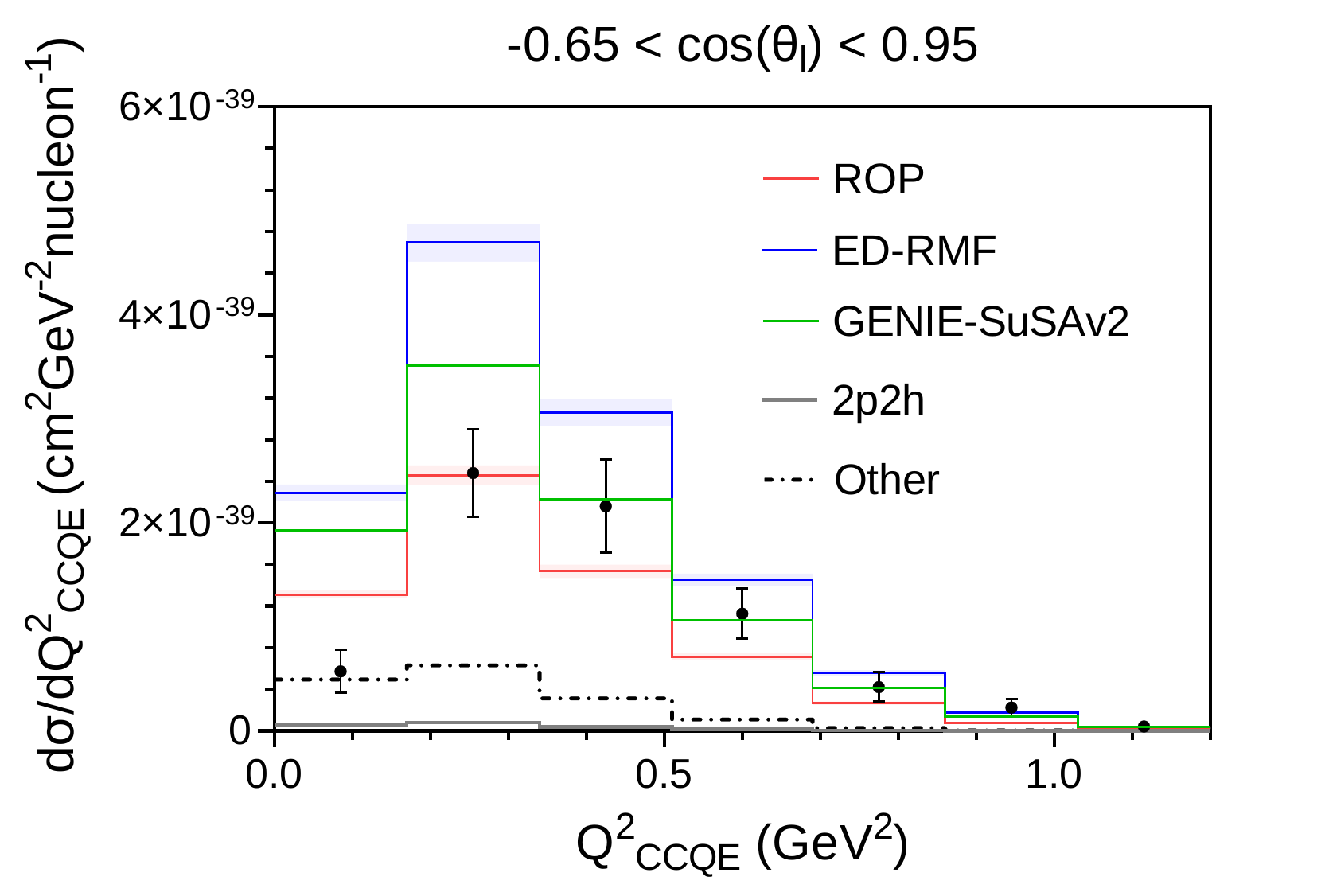} &\hspace{0.5cm}
			\includegraphics[width=0.32\textwidth]{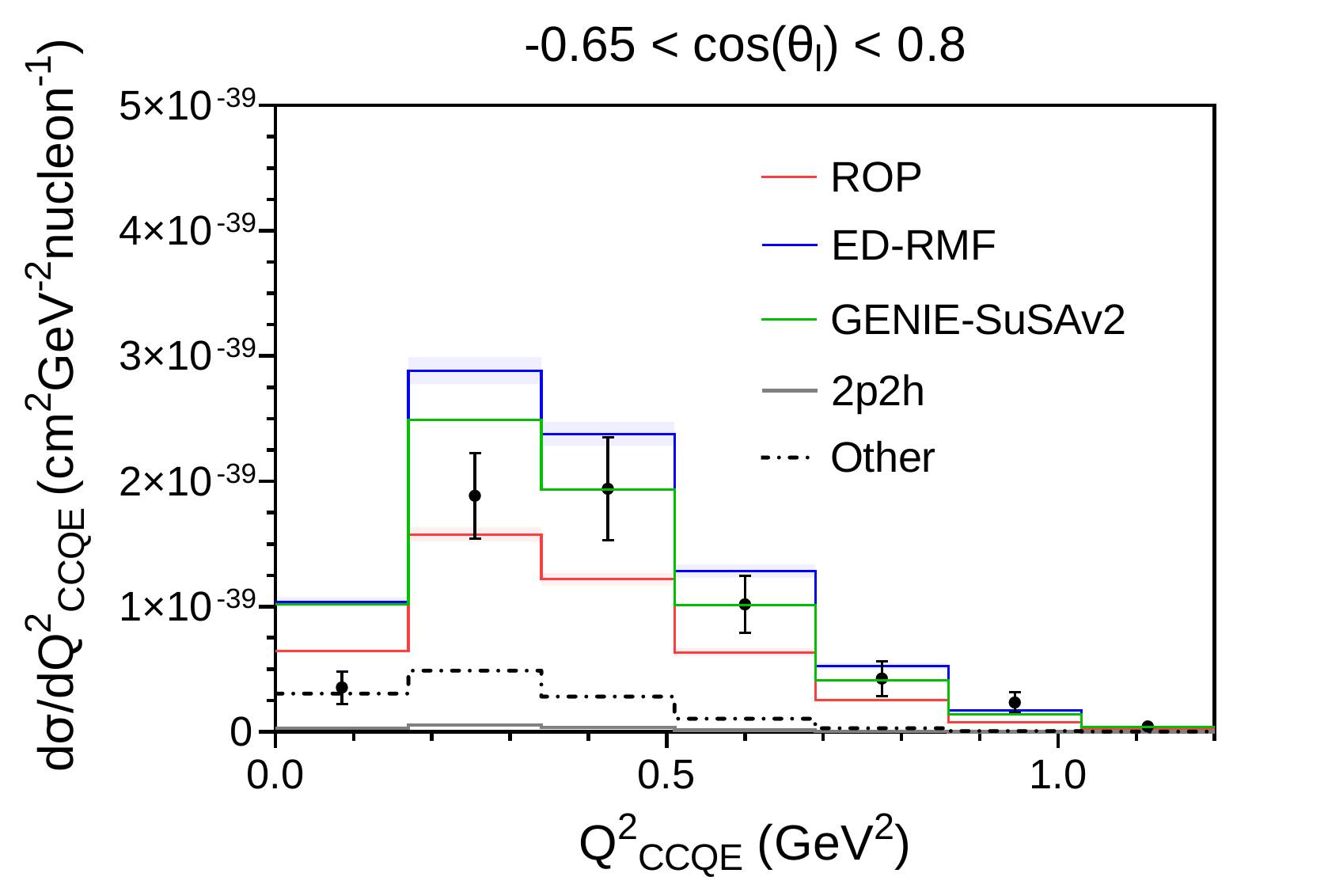} &\\ \includegraphics[width=0.32\textwidth]{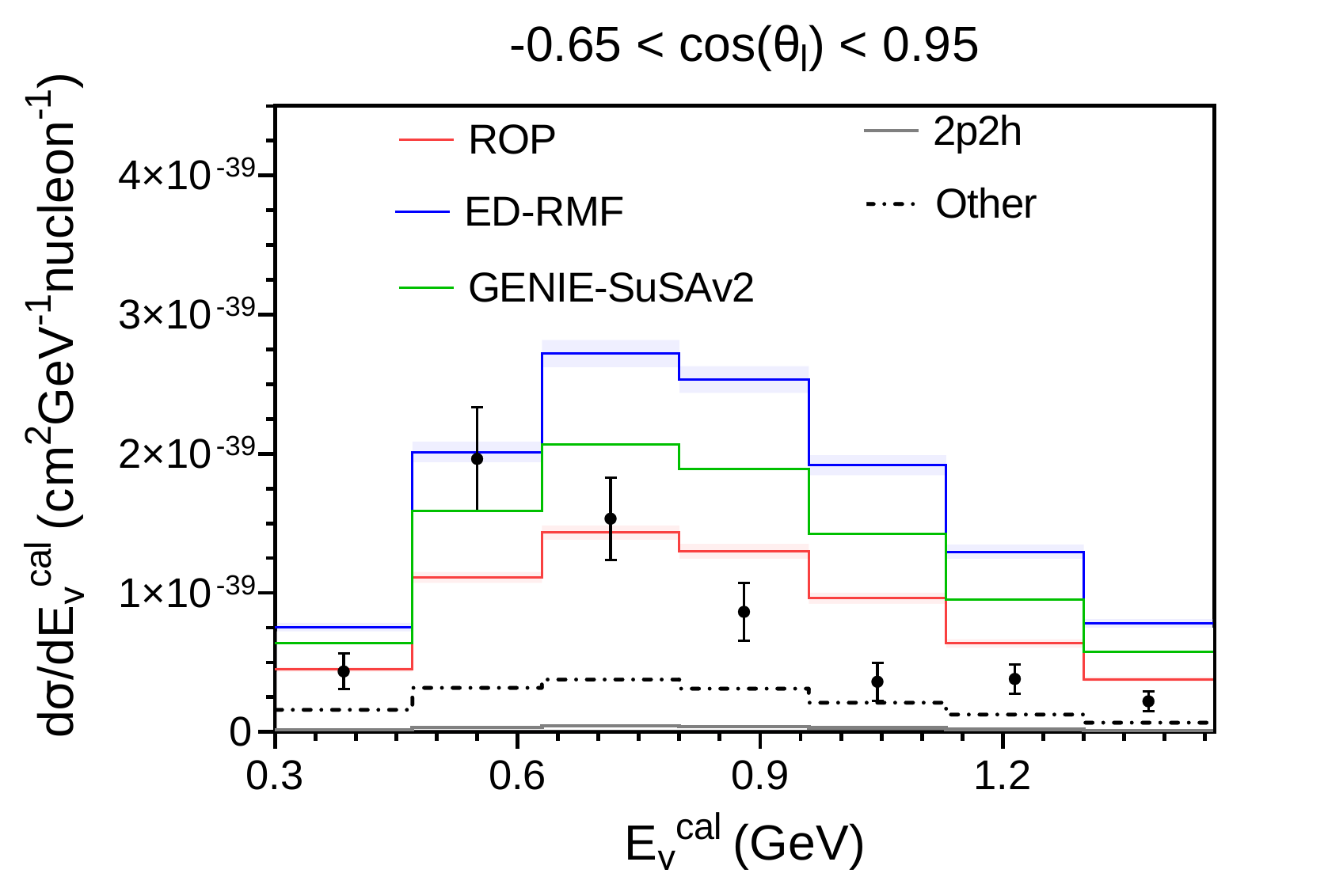} &\hspace{0.5cm}
			\includegraphics[width=0.32\textwidth]{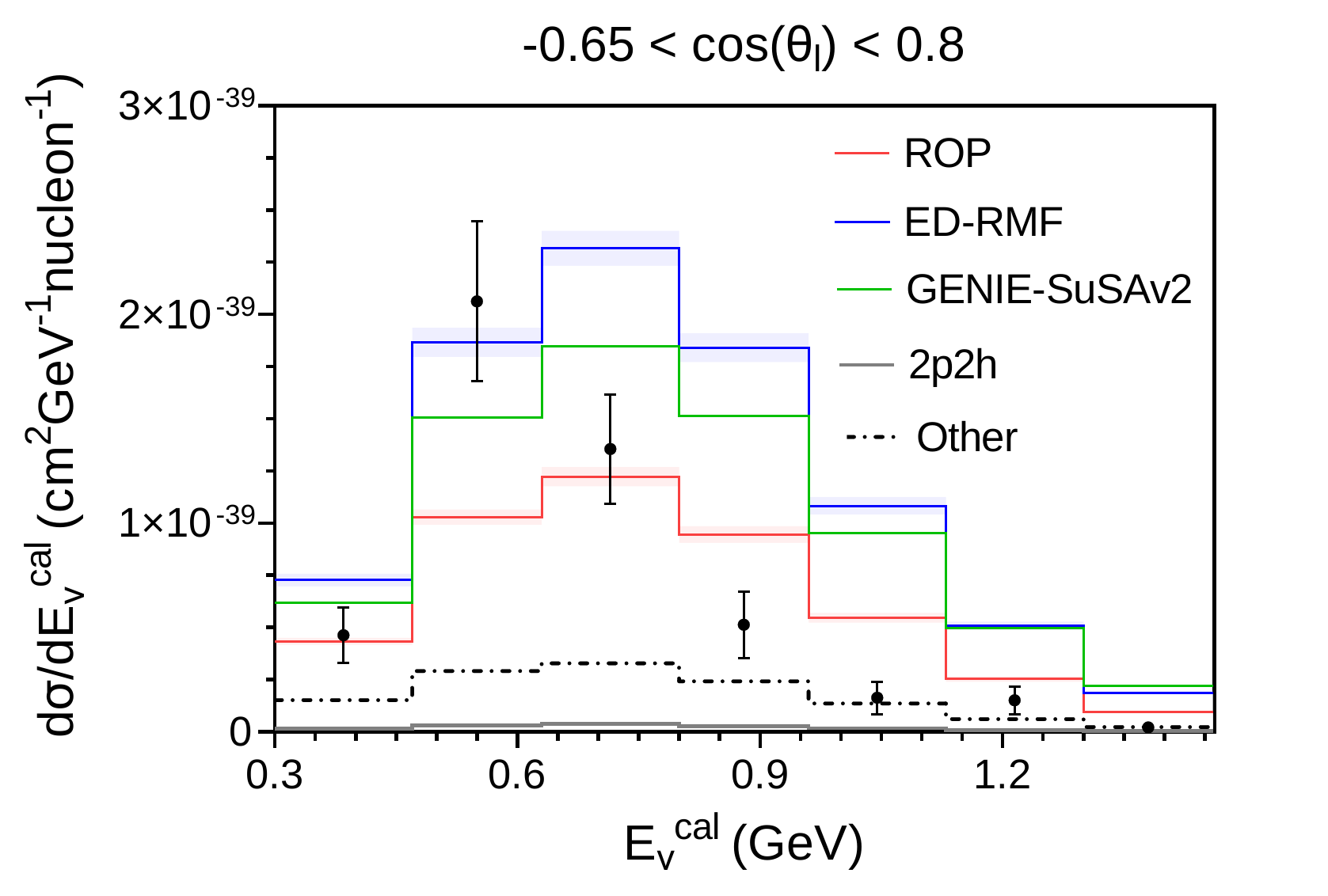} &\\	
		};
	\end{tikzpicture}
		\caption{\label{CC0pi1p}MicroBooNE CC$0\pi$1p $\nu_\mu$$-^{40}$Ar cross sections as function of the muon and proton momenta, proton polar scattering angle, $Q^2_{\textnormal{CCQE}}$ and $E_\nu^{\textnormal{cal}}$ as defined in Eq.~\ref{recons}. All curves include the two-particle-two-hole (denoted 2p2h) and pion absorption (denoted other) contributions evaluated using GENIE (shown separately). Experimental results were taken from~\cite{PhysRevLett.125.201803}. The bands drawn for the ED-RMF and ROP models are related with the uncertainties associated with the modeling of the initial nuclear state. The $\chi^2/\textit{d.o.f.}$ ratio is given in brackets in the legend of each distribution except for $Q^2_{\textnormal{CCQE}}$ and $E_\nu^{\textnormal{cal}}$ for which the covariance matrices are not available.}
\end{figure*}

\begin{figure}[!htbp]
	\centering
	\includegraphics[width=0.49\textwidth]{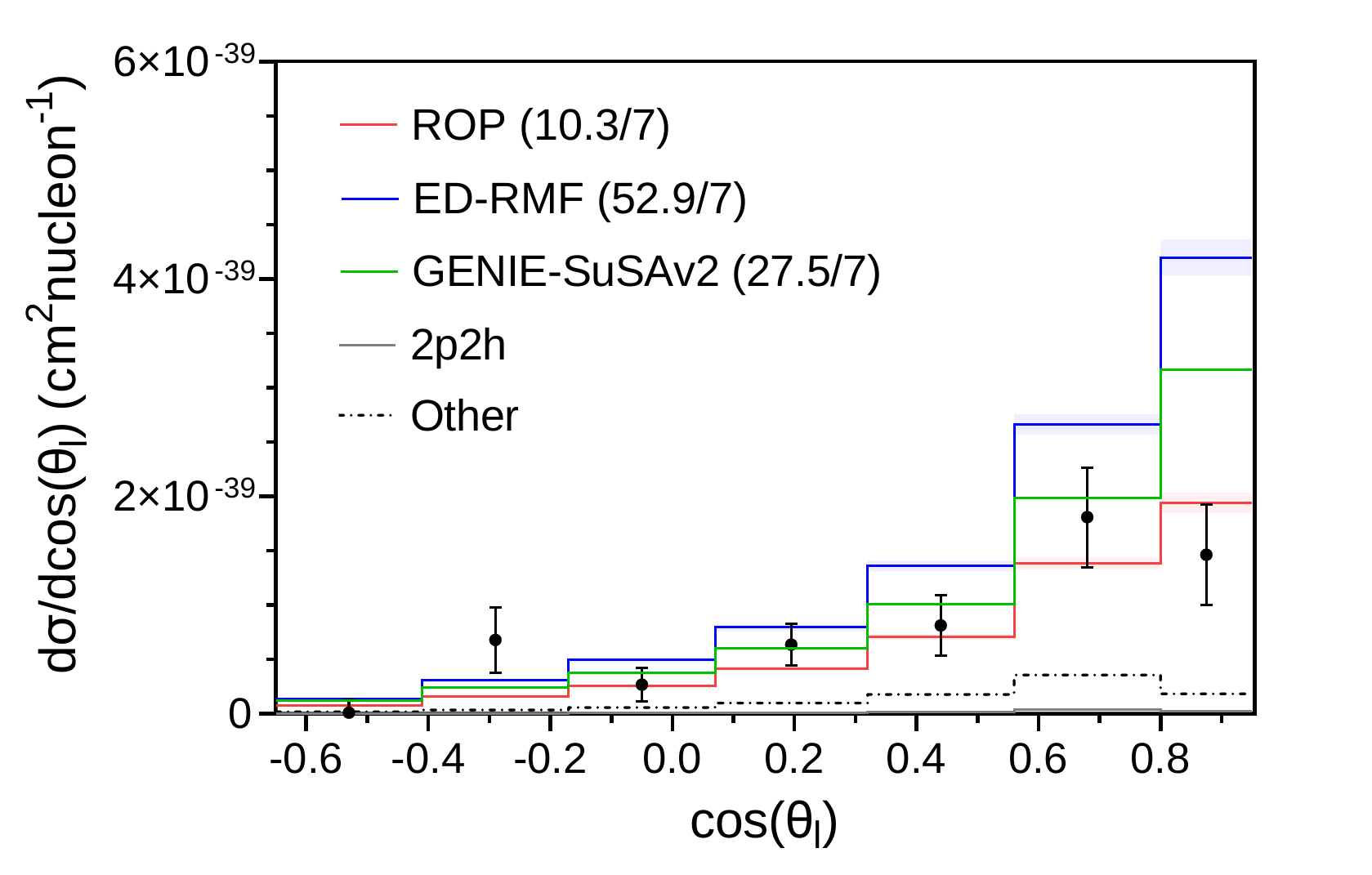}
	\caption{\label{CC0pi1p_costhetal}MicroBooNE CC$0\pi$1p $\nu_\mu$$-^{40}$Ar cross sections as function of the muon scattering angle. All curves include the two-particle-two-hole (denoted 2p2h) and pion absorption (denoted other) contributions evaluated using GENIE (shown separately). Experimental results were taken from~\cite{PhysRevLett.125.201803}. The bands drawn for the ED-RMF and ROP models are related with the uncertainties associated with the modeling of the initial nuclear state. The $\chi^2/\textit{d.o.f.}$ ratio is given in brackets in the legend of each distribution.}
\end{figure}

The MicroBooNE CC0$\pi$1p $\nu_\mu-^{40}$Ar measurements~\cite{PhysRevLett.125.201803} are shown in Figs.~\ref{CC0pi1p}~and~\ref{CC0pi1p_costhetal} as function of the muon and proton kinematics, together with the RDWIA and GENIE-SuSAv2 predictions. 
    
A noticeable difference with respect to the CC0$\pi$Np topology is the negligible contribution of the 2p2h channel in all distributions. The reason is the kinematic cuts applied to the CC0$\pi$1p signal which are summarized in Table~\ref{table: argon}. From a theoretical point of view, the CC0$\pi$1p topology, leaving aside the non-quasielastic contributions, is closer to the picture drawn by the ROP model, in which the imaginary part of the optical potential subtracts all the inelastic nuclear FSI, leaving only the elastic channel (i.e. the outgoing proton interacting elastically with the residual system). The inclusion of the background part of the spectral function introduces states with two or more nucleons being knocked out. Note that this contribution is very minor in the cross sections shown in Figs.~\ref{CC0pi1p}~and~\ref{CC0pi1p_costhetal}, and in fact, the result obtained after its subtraction is contained within the  uncertainty for both RDWIA predictions. The results presented as function of the proton kinematics show good agreement between the ROP and data, while the predictions by the other models overestimate the measured cross sections, especially the ED-RMF model. Regarding the lepton kinematics, the bins around the peak of the $k'$ distributions are slightly underestimated by the ROP but the ED-RMF and GENIE-SuSAv2 overestimate the data in the rest of the bins. 

In Fig.~\ref{CC0pi1p}, we also present the predictions as function of the reconstructed neutrino energy and $Q^2_\text{CCQE}$, which are defined as follows~\cite{PhysRevLett.125.201803}
	\begin{align}\label{recons}
		E_\nu^{\textnormal{cal}} = E_l + T_N + 40\;\textnormal{MeV}\, , \nonumber \\
		Q^2_{\textnormal{CCQE}} = \left(E_\nu^{\text{cal}}- E_l\right)^2 - 	\left(\mathbf{k}-\mathbf{k'}\right)^2
	\end{align}		
with $E_l$ the muon energy and $T_N$ the kinetic energy of the ejected proton. The argon binding energy is assumed to be $40$ MeV. Both RDWIA calculations tend to underestimate (ROP) or overestimate (ED-RMF) the measurements as function of $Q^2_\text{CCQE}$. However, the ED-RMF prediction describes better the data in the bin that excludes forward muon angles, i.e. $-0.65 <\cos{\theta_l}< 0.8$. In the case of the cross section as function of $E_\nu^\text{cal}$, all the models overpredict the data in the tail of the distribution at large $E_\nu^\text{cal}$-values. Given that a good description of $E_\nu^\text{cal}$ requires a description of the fully exclusive final state including very low momentum hadrons below detection threshold, the poor agreement is unsurprising.
    
Finally, the $\cos{\theta_l}$ distribution is shown in Fig.~\ref{CC0pi1p_costhetal}. The ROP and GENIE-SuSAv2 predictions are within the experimental uncertainty except for the forward angle bin that is overestimated by the GENIE-SuSAv2 and ED-RMF models. However, recent work by the MicroBooNE collaboration~\cite{https://doi.org/10.48550/arxiv.2301.03700} shows a $\cos{\theta_l}$ distribution that is reproduced correctly by different neutrino generators. This suggests that the discrepancy observed in Fig.~\ref{CC0pi1p_costhetal} at small muon scattering angles might be due to the use of an old version of the GENIE configuration that accounts for efficiency corrections and beam-induced backgrounds. This old version of GENIE (v2.12.2)~\cite{ANDREOPOULOS201087,andreopoulos2015genie} treats the nucleus as a Bodek-Ritchie Fermi gas, uses the Llewellyn-Smith CCQE scattering prescription~\cite{LLEWELLYNSMITH1972261}, the empirical MEC model~\cite{10.1063/1.4919465}, the Rein-Sehgal resonance and coherent scattering model~\cite{REIN198179}, and a data-driven FSI model denoted as “hA”~\cite{SGMashnik_2006}.

As already discussed, the ED-RMF prediction is an estimate of the `at least one proton in the final state' signal. In the case of GENIE-SuSAv2, the primary proton (the one at the neutrino interaction vertex) is introduced in the GENIE cascade, where it may undergo inelastic FSI leading to events with more than one proton (or pion creation) in the final state, which do not contribute to the signal. Therefore, in this CC0$\pi$1p case, ED-RMF is expected to be above GENIE-SuSAv2, and above the data. The fact that GENIE-SuSAv2 is systematically larger than ROP may be due to events in which the primary proton underwent inelastic FSI but they still contribute to the signal because, e.g., a undetectable neutron was knocked out, a second knocked out proton below threshold, etc. This observation could also be due to poor modeling of the elastic channel~\cite{nikolakopoulos2022benchmarking}, possibly or in part arising from the use of an inclusive model to predict semi-inclusive scenarios~\cite{psf2023008010}. Complete understanding requires further investigation.


\section{\label{sec4}Conclusions}

This paper presents a comparison of semi-inclusive $\nu_\mu-^{40}$Ar and $\nu_e-^{40}$Ar cross section measurements with two different theoretical approaches: RDWIA calculations using ED-RMF and ROP FSI models, and the SuSAv2 model implemented in GENIE. For the RDWIA models the calculation is performed taking into account conservative uncertainties associated with the modeling of the spectral function used for the description of the initial state.
 
Among the two RDWIA approaches considered in this work, the ROP model provides the best overall agreement with data for both CC0$\pi$Np and CC0$\pi$1p topologies. Only ROP is able to obtain a quantitatively reasonable agreement with the measurements, achieving an average $\chi^2$/\textit{d.o.f} across all measurements of 0.95 compared to 2.84 for ED-RMF. It is worth mentioning in the CC0$\pi$1p case the accordance between the ROP predictions and data as function of the muon and proton kinematics, except for the forward muon scattering angles (see Fig.~\ref{CC0pi1p_costhetal}) where the reported very low data point appears not to be present in subsequent analyses.

To conclude, the present study shows that the ROP model is in better agreement with data than the GENIE-SuSAv2 model for most of the kinematics explored. This comparison provides useful information on the kinematic regions where the GENIE-SuSAv2 results provide reasonable agreement with data and those where they are not reliable. The contributions from the 2p2h channel to the reaction analysed in this work, as well as future microscopic semi-inclusive 2p2h calculations, could be validated against additional exclusive measurements like CC0$\pi$ 1 muon and two protons in the final state topology measured by the MicroBooNE collaboration~\cite{microboonecollaboration2023measurement}.
\begin{acknowledgments}
	
This work is part of the I+D+i projects PID2020-114687GB-100; PID2021-127098NA-I00 funded by MCIN/AEI/10.13039/501100011033/FEDER, UE; PR65/19-22430 funded by the government of Madrid and Complutense University; and FQM160, SOMM17/61015/UGR and P20-01247 funded by Junta de Andaluc\'ia.
It is supported in part by the University of Tokyo ICRR's Inter-University Resarch Program FY2021 \& FY2022 (J.A.C., M.B.B., J.M.F.-P., G.D.M., R.G.-J.); by the European Union's Horizon 2020 research and innovation programme under the Marie Sklodowska-Curie grant agreement No. 839481 (G.D.M.); by the University of Turin, project BARM-RILO-21; and by INFN, national project NUCSYS (M.B.B. and J.M.F.-P.). J.M.F.-P. acknowledges support from a fellowship from the Ministerio de Ciencia, Innovaci\'on y Universidades, Program FPI (Spain).

\end{acknowledgments} 

\bibliography{references}
\end{document}